\documentclass[amsmath, aps]{revtex4}
\usepackage[pdftex]{graphicx}
\usepackage{times}
\usepackage{epsfig}
\usepackage{amssymb}

\usepackage{amsmath}
\begin{document}
\title{QED model of the radiation escape from the matter}
\author{T. Zalialiutdinov$^{1}$, D. Solovyev$^1$ and L. Labzowsky$^{1,2}$}

\affiliation{ 
$^1$ V. A. Fock Institute of Physics, St. Petersburg
State University, Petrodvorets, Oulianovskaya 1, 198504,
St. Petersburg, Russia
\\
$^2$  Petersburg Nuclear Physics Institute, 188300, Gatchina, St.
Petersburg, Russia}
\begin{abstract}
A simple model based on QED is presented for the estimation of contribution of the excited level few-photon decays to the radiation escape from the matter in the epoch of the cosmological hydrogen recombination. It is shown that apart from the widely studied two-photon decays, some specific 3-photon decays can contribute on the level of $0.1\%$ accuracy, required by the recent astrophysical observations.
\end{abstract}
\maketitle

\section{Introduction}

Theory of the cosmological hydrogen recombination became one of the most intensively discussed fundamental physics problems in the last decade. The interest comes from the accurate measurements of the asymmetry in the temperature and polarization distribution of the Cosmic Microwave Background (CMB) \cite{Hin}, \cite{Page}. The launching of the Planck Surveuor which enables to perform these measurements with accuracy $0.1\%$ makes the situation even more intriguing.

The modern theory of the cosmological hydrogen recombination starts from the papers by Zel'dovich, Kurt and Sunyaev \cite{Zeld} and by Peebles \cite{Peebles}. It was argued that the bound-bound one-photon transitions from the upper levels to the lower ones did not permit the hydrogen atoms to reach their ground states: each photon released in such transition in one atom was immediately absorbed by another atom. These reabsorption processes did not allow the radiation to "escape" the interaction with the matter. As it was first established in \cite{Zeld}, \cite{Peebles} the two-photon 2s-1s transition presents the main channel for the radiation "escape" and formation of the CMB. This transition also led to the final hydrogen recombination. Hence, the recent properties of the CMB are essentially defined by the two-photon processes during the cosmological recombination epoch.

In \cite{Dubrovich} the importance of the two-photon decays from excited states with $n>2$ for the detailed analysis of the properties of CMB was noted. Over the past few years the theory of cosmological recombination was essentially detalized by many authors. In particular, in \cite{Dubrovich}, \cite{Wong} it was demonstrated that the two-photon transitions $ns\rightarrow 1s(n>2)$ and $nd\rightarrow 1s$ can also give a sizeable contribution to the radiation "escape". There is a difference between the decay of $ns(n>2)$, $nd$ states and the decay of the $2s$ state. This difference is due to the presence of cascade transitions as the dominant decay channels in the cases of $ns(n>2)$ and $nd$ levels. For the $2s$ level the cascade transitions are absent. The cascade photons can be effectively reabsorbed and therefore the problem of separation of the "pure" two-photon emission from the cascade photons arises in connection with the "escape" probability. This problem was intensively discussed during the last decade \cite{J.Chluba}-\cite{Amaro}. As it was proved in \cite{LSP} the separation of the "pure" two-photon emission for the $ns(n>2)$ and $nd$ levels is an ambiguous procedure. First this ambiguity was established for the two-photon transitions with cascades in the highly charged ions \cite{LabShon}. To reach the level of accuracy $0.1\%$ for the theoretical description of the properties of the CMB many effects should be taken into account in the astrophysical equations describing the radiation "escape" process: consequences of the universe expansion, thermodynamical properties, induced radiation, processes of the electron scattering, Raman scattering etc. Detailed analysis of the various distortions of the resonant optical line spectra is also required \cite{ChlubSun}, including the nonresonant corrections \cite{Karsh}.

This very complicated construction requires a careful treatment of the basic principles which this construction is standing upon; these principles are given by the Quantum Electrodynamics (QED). In this paper we will analyse these principles and demonstrate that following them one can find some additional effects, small but probably sizeable at the level $0.1\%$. Our treatment will remain in the frames of QED applied to the free atoms in the field of photons; the astrophysical aspects will be restricted to the introduction of temperature (i.e. thermodynamical equilibrium). Actually we consider the model universe containing two atoms only. The first atom is in excited state and the second one in the ground state. The first atom emits the radiation and arrives in the ground state too. If this radiation is not absorbed by the second atom, both atoms appear to be in the ground state: recombination occurs and the radiation has "escaped" the interaction with the matter. This "escape" does not coincide with the definition adopted in astrophysics and has sense only within the frames of our model. We assume however that our model can reproduce correctly the relative role of the higher excited states compared to $ 2s $ state in the process of hydrogen recombination.

Our paper is organized as follows. In Section II we describe the process of the photon scattering (i.e. the process of scattering of the photons emitted by one atom, on another atom) in QED and apply this description to the rescattering of the Lyman-alpha photons. In Section III we investigate how the two-photon emission from the $ns$, $nd$ levels is absorbed in the one-photon transitions. This investigation will enable us to compare the probability of the radiation "escape" from the $ns(n>2)$, $nd$ levels with the "escape" from $2s$ level. Section IV is devoted to the studies of the multi-photon (i.e. 3-, 4-photon) transitions in the two-photon approximation. We will show that these transitions can give a non-negligible contribution to the radiation "escape". Section V contains discussion of the results and conclusions. In Appendix A we give a rigorous QED derivation of the Lorentz contour for the one-photon transition between the two arbitrary excited levels; such a derivation was yet absent in the literature. Appendix B is devoted to the derivation of the basic formula employed in section III.

\section{QED theory for the photon rescattering on an atom}
\subsection{Emission line profile for the transitions between two arbitrary levels}

A quantum mechanical phenomenological description of the line profile (Lorentz profile) is known since 1930 \cite{WW}, \cite{Heitler}. A QED derivation of the Lorentz profile for the transition between two excited states was given in \cite{LabShon}, \cite{AndrLab}. The corresponding expression for the transition $a\rightarrow a_1$ looks like \cite{AndrLab}:
\begin{eqnarray}
\label{1}
dW_{aa_1(a_0)}(\omega)=\frac{1}{2\pi}\frac{\Gamma_{aa_1}\Gamma_{a_1a_0}(\Gamma_a+\Gamma_{a_1})}{\Gamma_a\Gamma_{a_1}}\frac{d\omega}{(\omega-\tilde{\omega}_{aa_1})^2+\frac{1}{4}(\Gamma_a+\Gamma_{a_1})^2}\;.
\end{eqnarray}

In Eq. (\ref{1}) it is assumed that the lower level $a_1$ decays in turn to the ground state $a_0$ via one-photon decay. Here $\Gamma_a$, $\Gamma_{a_1}$ are the total widths of the levels $a$, $a_1$ and $\Gamma_{aa_1}$, $\Gamma_{a_1a_0}$ are the partial widths corresponding to the transitions $a\rightarrow a_1$ and $a_1\rightarrow a_0$. Emission probabilities are connected with the partial widths via equalities
\begin{eqnarray}
\label{2}
W_{aa_1}=\Gamma_{aa_1};\qquad W_{a_1a_0}=\Gamma_{a_1a_0}\;.
\end{eqnarray}
Finally, $\tilde{\omega}_{aa_1}=E_a+L_a-E_{a_1}-L_{a_1}$, where $E_a$, $E_{a_1}$ are the one-electron energies and $L_a$, $L_{a_1}$ represent the Lamb shifts of the levels $a$ and $a_1$. Thus, in principle, the line profile for the transition $a\rightarrow a_1$ depends on the further decay channel for the lower state $a_1$. Actually this is the dependence on the branching ratios $b_{aa_1}=\Gamma_{aa_1}/\Gamma_a$ and $b_{a_1a_0}=\Gamma_{a_1a_0}/\Gamma_{a_1}$. In the simplest case when both levels $a$ and $a_1$ have only one decay channel $b_{aa_1}=b_{a_1a_0}=1$ Eq. (\ref{1}) simplifies to
\begin{eqnarray}
\label{3}
dW_{aa_1}(\omega)=\frac{1}{2\pi}\frac{(\Gamma_a+\Gamma_{a_1})d\omega}{(\omega-\tilde{\omega}_{aa_1})^2+\frac{1}{4}(\Gamma_a+\Gamma_{a_1})^2}\;.
\end{eqnarray}
Here the dependence on the state $a_0$ disappeared totally. In Appendix A we present also the derivation of the Lorentz profile for the most general case when the lower level $a_1$ decays not directly to the ground state $a_0$ but to the intermediate state $a_2$, then to the lower intermediate state $ a_3 $ and so on. The total chain of decays (the cascade) is $a\rightarrow a_1\rightarrow a_1\rightarrow a_3\rightarrow...\rightarrow a_0$. It is assumed that these decays are of one-photon type.

An important question is: how far from the resonance the wings of the Lorentz profile can be extended? The answer depends on importance of the so called nonresonant (NR) corrections which distort, in principle, the line profile. The NR corrections were first introduced in \cite{Low} and recently discussed in connection with atomic laboratory experiments in \cite{LSPS}-\cite{LSS}. In the astrophysical aspect the role of the NR corrections was studied in \cite{Karsh}. According to these studies, the Lorentz profile can be extended far from the resonance (actually, to infinity) without any serious errors. We will use this extension throughout this paper.

Employing the extension of the profiles discussed above we choose the normalization condition for the Lorentz profile as:
\begin{eqnarray}
\label{4}
dW(\omega)=L(\omega)d\omega\;,
\end{eqnarray}
\begin{eqnarray}
\label{5}
\int\limits_0^{\infty}L_{aa_1}(\omega)d\omega=1\;,
\end{eqnarray}
in case of Eq. (\ref{3}) and
\begin{eqnarray}
\label{6}
\int\limits_0^{\infty}L_{aa_1(a_0)}(\omega)d\omega=b_{aa_1}b_{a_1a_0}\;,
\end{eqnarray}
in case of Eq. (\ref{1}). Eqs (\ref{5}), (\ref{6}) represent the absolute probability for the photon to be emitted via the transition $a\rightarrow a_1$ with any frequency value. If there are no other decay channels apart from $a\rightarrow a_1$, this probability equals to 1. If such decay channels exist both for $a$ and $a_1$ levels, this probability is defined by the product of the branching ratios $b_{aa_1}b_{a_1a_0}$.

\subsection{Reemission of the photons emitted by one atom by another atom in the same transition.}

Now we assume that the radiation emitted in the transition $a\rightarrow a_1$ and having the frequency distribution defined by Eq. (\ref{3}), is absorbed by another atom via transition $a_1\rightarrow a$. The simplified form of the Lorentz profile takes place, in particular, for the most important for the cosmological recombination Lyman-alpha line. The absorption Lorentz profile is defined by the Eq. (\ref{3}) and the absorption probabilities are connected with partial widths via
\begin{eqnarray}
\label{7}
W_{a_1a}=\frac{g_{a_1}}{g_a}\Gamma_{aa_1},
\end{eqnarray}
where $g_{a_1}$, $g_a$ are the degeneracies for the states $a_1$, $a$. If the absorbed photons originate from the emission line $a\rightarrow a_1$ of another atom, the probability of the absorption and hence reemission of these photons should be defined as
\begin{eqnarray}
\label{8}
X^{(2)}_{aa_1}=\int\limits_0^{\infty}I_{aa_1}(\omega)L_{aa_1}(\omega)d\omega\;,
\end{eqnarray}
where $I_{aa_1}(\omega)=(\Gamma_a+\Gamma_{a_1})L_{aa_1}(\omega)$ is the dimensionless distribution of the incident photons. This function is normalized according to the condition
\begin{eqnarray}
\label{9}
\int\limits_0^{\infty}I_{aa_1}(\omega)d\omega=\Gamma_a+\Gamma_{a_1}\;.
\end{eqnarray}
The formal proof of Eq. (\ref{8}) on the basis of QED is given in Appendix B.\\
The frequency distribution of the emitted photons was first introduced in the QED S-matrix theory in \cite{Low} and later employed in \cite{LabSol}, \cite{LSPAS} for the studies of the multiple photon scattering on the hydrogen atom.

In this way we can define also the probability of the photon emission after the multiple (n-fold) scattering:
\begin{eqnarray}
\label{10}
X_{aa_1}^{(n)}=\int\limits_0^{\infty}\left[I_{aa_1}(\omega)\right]^{n-1}L_{aa_1}(\omega)d\omega=(\Gamma_a+\Gamma_{a_1})^{n-1}\int\limits_0^{\infty}\left[L_{aa_1}(\omega)\right]^{n}d\omega\;.
\end{eqnarray}
For $n=1$ the expression (\ref{10}) reduces to Eq. (\ref{5}). The integral can be extended to the interval $-\infty\leq \omega \leq +\infty$ since the main contribution comes from the pole in the expressions (\ref{1}) or (\ref{3}) for $L_{aa_1}$. Then we can evaluate the integral in the complex plane. Employing the equality
\begin{eqnarray}
\label{11}
\left[\frac{1}{(\omega-\tilde{\omega}_{aa_1})^2+\frac{1}{4}(\Gamma_a+\Gamma_{a_1})^2}\right]^n=\frac{1}{\left[\omega-\tilde{\omega}_{aa_1}-\frac{i}{2}(\Gamma_a+\Gamma_{a_1})\right]^n\left[\omega-\tilde{\omega}_{aa_1}+\frac{i}{2}(\Gamma_a+\Gamma_{a_1})\right]^n}
\end{eqnarray}
and using the Cauchy's formula we find
\begin{eqnarray}
\label{12}
X^{(n)}_{aa_1}=\left[(\Gamma_a+\Gamma_{a'})\right]^{n-1}\frac{2\pi i}{(n-1)!}\left[\Biggr(\frac{1}{\omega-\tilde{\omega}_{aa_1}+\frac{i}{2}(\Gamma_a+\Gamma_{a_1})}\Biggl)^n\right]^{(n-1)}_{\omega=\tilde{\omega}_{aa_1}+\frac{i}{2}(\Gamma_a+\Gamma_{a_1})}\times\left(\frac{\Gamma_a+\Gamma_{a_1}}{2\pi}\right)^n\;,
\end{eqnarray}
where $[...]^{(n-1)}$ denotes the $(n-1)$-fold derivative with respect to the variable $\omega$. The evaluation in Eq. (\ref{12}) results
\begin{eqnarray}
\label{13}
X^{(n)}_{aa_1}=\frac{(2n-2)!}{((n-1)!)^2}\frac{1}{(2\pi)^{n-1}}.
\end{eqnarray}
For all $n>1$ it is easy to check that $X^{(n)}_{aa_1}<1$, so that we can interprete $X^{(n)}_{aa_1}$ as the absolute probability of the absorption in the process of rescattering and reemission. Then the quantity
\begin{eqnarray}
\label{14}
Y^{(n)}_{aa_1}=1-X^{(n)}_{aa_1}=1- \frac{(2n-2)!}{((n-1)!)^2}\frac{1}{(2\pi)^{n-1}}\;.
\end{eqnarray}
can be interpreted as the probability of the radiation "escape". In particular,
\begin{eqnarray}
\label{15}
Y^{(2)}_{aa_1}=1-\frac{1}{\pi}=0.682\;,
\end{eqnarray}
\begin{eqnarray}
\label{16}
Y^{(3)}_{aa_1}=1-\frac{3}{2\pi^2}=0.848\;.
\end{eqnarray}
So, in our simple model the probability of the radiation "escape" directly via the arbitrary one-photon transition is $0.682$ already after the first rescattering and becomes close to 1 with the increase of number of rescatterings. This result does not depend on the particular transition and corresponds also to the Lyman-alpha transition.
We should stress that these estimates cannot replace the accurate astrophysical approach to the problem of the photon rescattering on the matter and are presented here only to make the further derivations more obvious.

\section{Radiation "escape" in the two-photon transitions}
\subsection{QED theory of the two-photon transitions}

Quantum mechanical theory for the two-photon transitions was first developed by G\"{o}ppert-Mayer \cite{Goepp} and the first evaluation of the two-photon $2s\rightarrow 1s+2\gamma(E1)$ decay rate in hydrogen was performed by Breit and Teller \cite{Breit}. The accurate nonrelativistic calculation for this transition rate was given in \cite{Klarsfeld}. The fully relativistic calculations valid also for the H-like ions with arbitrary nuclear charge $1\leq Z\leq 100$ were performed in \cite{GD82}-\cite{Santos}. The most accurate recent calculation with QED corrections one can find in \cite{Jent}.

The modifications of the theory necessary to describe the two-photon transitions with cascades were discussed in \cite{LabShon} (see also \cite{LSP}, \cite{AndrLab}).

The transition rate for the $2s\rightarrow 1s+2\gamma(E1)$ transition in H atom looks like (in atomic units)
\begin{eqnarray}
\label{17}
dW^{(2\gamma)}_{2s,1s}(\omega)=\frac{8\omega^3(\omega_0-\omega)^3}{27\pi}\alpha^2\left|S_{2s,1s}(\omega)+S_{2s,1s}(\omega_0-\omega)\right|^2d\omega\;,
\end{eqnarray}
where
\begin{eqnarray}
\label{18}
S_{2s,1s}(\omega)=\sum\limits_{n'p}\frac{<R_{1s}|r|R_{n'p}><R_{n'p}|r|R_{2s}>}{E_{n'p}-E_{2s}+\omega}\;,
\end{eqnarray}
\begin{eqnarray}
\label{19}
<R_{n'l'}|r|R_{nl}>=\int\limits_0^{\infty}r^3R_{n'l'}(r)R_{nl}(r)dr\;,
\end{eqnarray}
$\omega_0=E_{2s}-E_{1s}$, $R_{nl}(r)$ is the radial part of the nonrelativistic hydrogen wave function, $E_{nl}$ are the electron energies for the hydrogen atom and $\alpha$ is the fine structure constant. Due to the absence of the energy levels between $2s$ and $1s$ (i.e. cascades) the denominator in Eq. (\ref{18}) has no zeros.

The total decay rate for the two-photon transition $2s\rightarrow 1s$ can be obtained by integration of Eq. (\ref{17}) over the entire frequency interval
\begin{eqnarray}
\label{20}
W_{2s,1s}=\frac{1}{2}\int\limits_0^{\omega_0}dW_{2s,1s}=8.229\,\,s^{-1}\;.
\end{eqnarray}

An expression for the transition rate $W_{3s,1s}$ in the presence of cascades was given in \cite{LSP}:
\begin{eqnarray}
\label{21}
W_{3s,1s}^{(2\gamma)}=W_{3s,1s}^{(cascade)}+W_{3s,1s}^{(pure 2\gamma)}+W_{3s,1s}^{(interference)}\;,
\end{eqnarray}
where
\begin{eqnarray}
\label{22}
W^{({cascade})}_{3s;1s}=\frac{4}{27\pi}\frac{\Gamma_{3s}+\Gamma_{2p}}{\Gamma_{2p}}\int\limits_{({\bf  II})}\omega^3(\omega_0-\omega)^3\left|\frac{\langle R_{3s}(r)|r|R_{2p}(r)\rangle\langle R_{2p}(r')|r'|R_{1s}(r')\rangle}{E_{2p}-E_{3s}+\omega-\frac{i}{2}(\Gamma_{3s}+\Gamma_{2p})}\right|^2d\omega + 
\\
\nonumber
+\frac{4}{27\pi}\int\limits_{({\bf  IV})}\omega^3(\omega_0-\omega)^3\left|\frac{\langle R_{3s}(r)|r|R_{2p}(r)\rangle\langle R_{2p}(r')|r'|R_{1s}(r')\rangle}{E_{2p}-E_{1s}-\omega-\frac{i}{2}\Gamma_{2p}}\right|^2 d\omega\;,
\end{eqnarray}
\begin{eqnarray}
\label{23}
W_{3s;1s}^{(pure 2\gamma)}=\frac{4}{27\pi}\int\limits_{({\bf  II})}\omega^3(\omega_0-\omega)^3\left|S_{1s;3s}^{(2p)}(\omega)+S_{1s;3s}(\omega_0-\omega)\right|^2d\omega +
\nonumber
\\
+\frac{4}{27\pi}\int\limits_{({\bf  IV})}\omega^3(\omega_0-\omega)^3\left|S_{1s;3s}(\omega)+S_{1s;3s}^{(2p)}(\omega_0-\omega)\right|^2d\omega+
\\
\nonumber
+\frac{4}{27\pi}\int\limits_{({\bf  I+III+V})}\omega^3(\omega_0-\omega)^3\left|S_{1s;3s}(\omega)+S_{1s;3s}(\omega_0-\omega)\right|^2d\omega\;,
\end{eqnarray}
\begin{eqnarray}
\label{24}
dW^{(interference)}_{3s;1s}=\int\limits_{({\bf  II})}\frac{4\omega^3(\omega_0-\omega)^3}{27\pi}Re\left[\frac{\langle R_{3s}(r)|r|R_{2p}(2r)\rangle\langle R_{2p}(r')|r'|R_{1s}(r')\rangle}{E_{2p}-E_{3s}+\omega-\frac{i}{2}\Gamma_{2p}}\right]\left[S_{1s;3s}^{(2p)}(\omega)+S_{1s;3s}(\omega_0-\omega)\right]d\omega+
\nonumber
\\
+\int\limits_{({\bf  IV})}\frac{4\omega^3(\omega_0-\omega)^3}{27\pi}Re\left[\frac{\langle R_{3s}(r)|r|R_{2p}(2r)\rangle\langle R_{2p}(r')|r'|R_{1s}(r')\rangle}{E_{2p}-E_{1s}-\omega-\frac{i}{2}\Gamma_{2p}}\right]\left[S_{1s;3s}(\omega)+S_{1s;3s}^{(2p)}(\omega_0-\omega)\right]d\omega\qquad\;.
\end{eqnarray}
Here $S_{1s;3s}^{(2p)}(\omega)$ is the expression (\ref{18}) with the term in the $n'p=2p$ excluded from the summation, $\omega_0=E_{3s}-E_{1s}$.
The intervals of integration over $\omega$ {\bf (I)-(V)} are defined as
\begin{eqnarray}
\label{25}
{\bf (I)} \quad 	 0\leq \omega\leq \omega_{01}-l(\Gamma_{2p}+\Gamma_{3s})\;,
\end{eqnarray}
\begin{eqnarray}
\label{26}
{\bf (II)} \quad 	 \omega_{01}-l(\Gamma_{2p}+\Gamma_{3s})\leq \omega\leq \omega_{01}+l(\Gamma_{2p}+\Gamma_{3s})\;,
\end{eqnarray}
\begin{eqnarray}
\label{27}
{\bf (III)} \quad 	 \omega_{01}+l(\Gamma_{2p}+\Gamma_{3s})\leq \omega\leq \omega_{02}-l(\Gamma_{2p}+\Gamma_{3s})\;,
\end{eqnarray}
\begin{eqnarray}
\label{28}
{\bf (IV)} \quad 	 \omega_{02}-l(\Gamma_{2p}+\Gamma_{3s})\leq \omega\leq \omega_{02}+l(\Gamma_{2p}+\Gamma_{3s})\;,
\end{eqnarray}
\begin{eqnarray}
\label{29}
{\bf (V)} \quad 	 \omega_{02}+l(\Gamma_{2p}+\Gamma_{3s})\leq \omega\leq \omega_{0}\;,
\end{eqnarray}
where $\omega_{01}=E_{3s}-E_{2p}$, $\omega_{02}=E_{2p}-E_{1s}$ are the frequencies for the two links of the cascade, $l$ is integer chosen to separate the cascade contribution from the "pure" two-photon contribution. As it was shown in \cite{LSP}, this separation is not unique, i.e. the contributions $W^{(cascade)}_{3s,1s}$, $W^{(pure 2\gamma)}_{3s,1s}$, $W^{(interference)}_{3s,1s}$ change essentially depending on the choice of the $l$, but the total sum $W^{(2\gamma)}_{3s,1s}$ remains invariant: 
\begin{eqnarray}
W^{(2\gamma)}_{3s,1s}=\frac{1}{2}\int\limits_0^{\omega_0}dW_{3s,1s}d\omega=6.317\cdot 10^{6}\;.
\end{eqnarray}
Note that the factor $ \frac{\Gamma_{3s}+\Gamma_{2p}}{\Gamma_{2p}} $ in the first line in Eq. (22) was omitted in \cite{LSP}, so the numerical value Eq. (30) also was different.
We should stress, that in our derivations the total width of the level $\Gamma_{3s}$ does not coincide with the value given by the Eq. (30), but coincides with the transition rate $ \Gamma^{(1\gamma)}_{3s,2p} $,as in the atomic spectroscopy. The value $\Gamma^{(1\gamma)}_{3s,2p}$, in principle, defines the total width in the laboratory experiments when the one-photon transition rate for the photons with frequency $\omega_{3s,2p}=E_{3s}-E_{2p}$ is measured. The one-photon decay $3s\rightarrow 2p$ appears to be faster than the decay $3s\rightarrow 2p\rightarrow 1s$ due to the destructive interference of the cascade decay with the "pure"\; two-photon decay in Eq. (30). However this difference can be traced only in the fifth digit. The same picture holds for the decays of the other $ns(n>2)$, $nd$ levels. Similar expressions can be written for the transition $3d-1s$ with the cascade $3d-2p-1s$, for the transition $4s-1s$ with two cascades $4s-3p-1s$ and $4s-2p-1s$, etc.

\subsection{Radiation "escape" via two photon decays}

We define the "escape" probability for the incoming two-photon radiation via the Lyman-alpha channel similarly to Eq. (\ref{8}):
\begin{eqnarray}
\label{31}
X^{(2)2\gamma}_{2s,1s}=\frac{1}{2}\int\limits_0^{\omega_0}I_{2s,1s}(\omega)L_{2p,1s}(\omega)d\omega\;,
\end{eqnarray}
where $I_{2s,1s}(\omega)=dW_{2s,1s}(\omega)$, $\omega_0=E_{2s}-E_{1s}$. The result of the integration
\begin{eqnarray}
\label{32}
X^{(2)2\gamma}_{2s,1s}=6.50\cdot 10^{-22}
\end{eqnarray}
shows that the two-photon $2s-1s$ radiation emitted by one atom cannot be absorbed by another atom. This means that the radiation "escape" via the two-photon $2s-1s$ transition is absolutely full:
\begin{eqnarray}
\label{33}
Y^{(2)2\gamma}_{2s,1s}=1-X^{(2)2\gamma}_{2s,1s}=1\;.
\end{eqnarray}
The superscript $(2)$ here, as in Section II, means that we consider only one scattering (reemission) of the photons. This is enough to understand the relative importance of different decay channels for the radiation "escape".

Now we can repeat the same for the transition $3s\rightarrow 1s+2\gamma(E1)$. In this case we evaluate the probability of reemission of the $3s-1s$ two-photon radiation via all possible one-photon transitions within the frequency range $ [0,\omega_0]$, i.e.  $3s-2p$, $3d-2s$, $3p-2s$ and $2p-1s$
\begin{eqnarray}
\label{34}
X^{(2)2\gamma}_{3s,1s}=\frac{1}{2}\int\limits_0^{\omega_0}I_{3s,1s}(\omega)\left[L_{3s,2p}(\omega)+L_{2p,1s}(\omega)+L_{3d,2p}(\omega)+L_{3p,2s}(\omega)\right]d\omega\;,
\end{eqnarray}
where $\omega_0=E_{3s}-E_{1s}$.The numerical result is
\begin{eqnarray}
\label{35}
X^{(2)2\gamma}_{3s,1s}=0.00497\;.
\end{eqnarray}
The value $X^{(2)2\gamma}_{3s,1s}$ is much larger than $X^{(2)2\gamma}_{2s,1s}$ but still essentially smaller than 1. This means that the "escape" probability is very high:
\begin{eqnarray}
\label{36}
Y^{(2)2\gamma}_{3s,1s}=0.99504\;.
\end{eqnarray}

The same picture holds for the two-photon decays of the other $ns$ $(n>2)$, $nd$ levels: for the transition 
$3d\rightarrow1s+2\gamma(E1)$ which occurs as one cascade with two links; for the transition $4s\rightarrow1s+2\gamma(E1)$ which includes two cascades each with two links $4s\rightarrow3p\rightarrow1s$ and $4s\rightarrow2p\rightarrow1s$  and for the transition $4d\rightarrow1s+2\gamma(E1)$ which includes two cascades each with two links  $4d\rightarrow3p\rightarrow1s$ and $4d\rightarrow2p\rightarrow1s$. The corresponding total decay rates $W^{(2\gamma)}_{3d,1s}$, $W^{(2\gamma)}_{4s,1s}$, $W^{(2\gamma)}_{4d,1s}$ and the total widths of the levels $\Gamma_{3d}$, $\Gamma_{4s}$, $\Gamma_{4d}$ as well as the probabilities of the reemission $X^{(2)2\gamma}_{nl,1s}$ and the "escape" probabilities $Y^{(2)2\gamma}_{nl,1s}$ are given in Table 1.

Thus, all the levels $ns$, $nd$ with $n=3,4$ seem to be nearly as effective for the radiation "escape" via the two-photon transitions to the $1s$ state, as the $2s$ level. The smallest "escape" (difference about $3\%$ with $2s$ level) occurs for the $3d$ level. This corresponds to the maximum "death probability" for the Lyman-alpha photons due to the transitions to $3d$ level, as found in \cite{ChlubSun}. However, the role of all these levels in the cosmological radiation "escape" is strongly suppressed by the thermodynamical factor (see Section V).

Two comments are necessary concerning the accuracy of the results given above. First, we have fully neglected the two-photon transitions other than E1E1. For the neutral hydrogen atom it is well justified (see, for example, \cite{LSPSEur}). Second, we have neglected the difference between $\Gamma_{aa'}$ and $\Gamma_a$ in Eq. (\ref{1}) for the one-photon transitions $3s-2p$, $3d-2p$, $4s-3p$, $4s-2p$, $4d-3p$, $4d-2p$. For example, in Eq. (\ref{34}) the Lorentz profile $L_{3s,2p}$ should be defined as
\begin{eqnarray}
\label{38}
L_{3s,2p}=\frac{1}{2\pi}\frac{\Gamma^{(1\gamma)}_{3s,2p}}{\Gamma_{3s}}\frac{\Gamma_{3s}^{(1\gamma)}+\Gamma_{2p}^{(1\gamma)}}{(\omega-\tilde{\omega}_{3s,2p})^2+\frac{1}{4}(\Gamma_{3s}^{(1\gamma)}+\Gamma_{2p}^{(1\gamma)})^2}\;.
\end{eqnarray}

As we have discussed above in our derivations we have to put $\Gamma_{3s}=\Gamma^{(1\gamma)}_{3s,2p}$. In the latter equality the transitions $3s\rightarrow 1s+\gamma(M1)$, $3s\rightarrow 2s+\gamma(M1)$ are neglected, which give only extremely small contribution to the total width $\Gamma_{3s}$ \cite{Sucher}.

\section{Radiation "escape" in the multiphoton transitions}
\subsection{Contribution of the 3-photon transitions}

In \cite{SL}, \cite{SLCan} it was suggested that the multiphoton transitions which contain cascades with the two-photon links can also contribute to the radiation "escape" in the process of the cosmological recombination. This approach was called "two-photon approximation" since the contribution of the "pure" multiphoton transitions with the number of photons more than two were neglected. One of the examples described in \cite{SL} was the two-photon approximation for the 3-photon $3p\rightarrow 1s$ transition. The $3p$ level decay can occur either as a one-photon transition $3p\rightarrow 1s+\gamma(E1)$ or as a 3-photon transition $3p\rightarrow 1s+3\gamma(E1)$. These channels do not interfere due to the different number of photons in the final state. The one-photon decay rate is
\begin{eqnarray}
\label{39}
W^{(1\gamma)}_{3p,1s}=195.61 m\alpha^2(\alpha Z)^4\,\,r.u. = 1.67342\cdot 10^8\,\, s^{-1}\;,
\end{eqnarray}
where $m$ is the electron mass, $\alpha$ is the fine structure constant, $Z$ is the charge of the nucleus ($Z=1$ for the hydrogen). The 3-photon decay rate $3p\rightarrow 1s+3\gamma(E1)$ consists of the "pure" 3-photon contribution, two cascade contributions $3p\rightarrow 2s+ \gamma(E1)\rightarrow 1s+2\gamma(E1)$, $3p\rightarrow 2p+ 2\gamma(E1)\rightarrow 1s+\gamma(E1)$ and the interference terms. The "pure" 3-photon contribution to the decay rate is of the order $m\alpha^3(\alpha Z)^8$ r.u. Three-photon contribution was evaluated in \cite{SLPSh} for the $2p\rightarrow  1s+3\gamma(E1)$ transition which does not contain any cascade contributions:
\begin{eqnarray}
\label{40}
W^{(3\gamma)}_{2p,1s}=0.4946 m\alpha^3(\alpha Z)^8\,\,r.u.
\end{eqnarray}
In principle, for $3p\rightarrow 1s+3\gamma(E1)$ transition rate the contributions of the "pure" 3-photon decay channel and the cascade contributions are again inseparable, similar to the case of the 2-photon $3s\rightarrow 1s+2\gamma(E1)$, $3d\rightarrow 1s+2\gamma(E1)$ transitions as discussed in Section III. However, unlike the two-photon decays in Section III, where at the level of accuracy of the "two-photon approximation" we were interested in the all contributions, in case of the 3-photon transitions at the same level of accuracy we have to keep only the cascade contributions and neglect fully the "pure" 3-photon contributions and the interference terms. This simplifies our task and retaining only the cascade terms, we find \cite{SL}
\begin{eqnarray}
\label{41}
W^{(3\gamma)}_{3p,1s}=W^{(2\gamma)}_{3p,2p}+\frac{W^{(1\gamma)}_{3p,2s}}{\Gamma_{3p}}W^{(2\gamma)}_{2s,1s}\;.
\end{eqnarray}
Note, that in \cite{SLPSh} the right-hand side of the equation corresponding to Eq. (40) contained the wrong factor $ \frac{3}{4} $, the same concerns the equation for the other two-photon decay rates. This mistake was noticed in \cite{answer1}.
The order of the magnitude of the 3-photon cascade transition rate is defined by the fact that the cascade transition rate is determined by transition rate of the slowest cascade link, i.e. in our case by the two-photon transitions. 

The total decay rate of the $3p$ level $\Gamma_{3p}$ is defined as (see discussion concerning the width $\Gamma_{3s}$ in Section II)
\begin{eqnarray}
\label{42}
\Gamma_{3p}=\Gamma_{3p,1s}^{(1\gamma)}+\Gamma_{3p,2s}^{(1\gamma)}\;.
\end{eqnarray}

The two-photon transition rate $W^{(2\gamma)}_{3p,2p}$ should be evaluated similarly to $W^{(2\gamma)}_{2s,1s}$ transition rate since it is "pure" two-photon transition rate. Hence
\begin{eqnarray}
\label{43}
W^{(2\gamma)}_{3p,2p}=\frac{1}{2}\int\limits_0^{\omega_0}dW^{(2\gamma)}_{3p,2p}(\omega)\;,
\end{eqnarray}
where $\omega_0=E_{3p}-E_{2p}$. The two-photon frequency distribution $dW^{(2\gamma)}_{3p,2p}\equiv I_{3p,2p}$ looks like (in a.u.)
\begin{eqnarray}
\label{44}
dW^{(2\gamma)}_{3p,2p}(\omega)=\frac{8\omega^3(\omega_0-\omega)^3}{9\pi \ast15^2}\alpha^2\sum\limits_{m_{l_{3p}}m_{l_{2p}}}\sum\limits_{q'q}(-1)^{q+q'}\left|S^{qq'}_{3p,2p}(\omega)+S^{q'q}_{3p,2p}(\omega_0-\omega)\right|^2d\omega\;,
\end{eqnarray}
where
\begin{eqnarray}
\label{45}
S^{q'q}_{3p,2p}(\omega)=(5C^{l_{3p} m_{l_{3p}}}_{1\; q'\; 0\; 0}C^{0 \;0}_{1\; q\; l_{2p} m_{l_{2p}}})\sum\limits_{n's}\frac{<R_{3p}|r|R_{n's}><R_{n's}|r|R_{2p}>}{E_{n's}-E_{3p}+\omega}+\nonumber\\+(2\sqrt{5}\sum\limits_{m_{l_n}}C^{l_{3p} m_{l_{3p}}}_{1\; q'\; 2\; m_{l_n}}C^{2\; m_{l_n}}_{1 \;q\; l_{2p} m_{l_{2p}}})\sum\limits_{n'd}\frac{<R_{3p}|r|R_{n'd}><R_{n'd}|r|R_{2p}>}{E_{n'd}-E_{3p}+\omega}\;,
\end{eqnarray}
\begin{eqnarray}
\label{46}
<R_{n'l'}|r|R_{nl}>=\int\limits_0^{\infty}r^3R_{n'l'}(r)R_{nl}(r)dr\;,
\end{eqnarray}
$C^{l\; m_{l}}_{l_1\; m_{l_1}\; l_2\; m_{l_2}}$ are the Clebsh-Gordan coefficients, $ m_l $ are the angular momentum projections.\\
\\
Unlike Eq.(\ref{17}) there are no resonant terms with zero denominators in Eq.(\ref{45}). Therefore we can perform the summation over $n's$, $n'd$ in Eq.(\ref{45}) explicitly using the known expressions  for the Coulomb Green function \cite{RZM}. Performing also the summation over all angular momentum projections in Eq. (\ref{45}), we arrive at
\begin{eqnarray}
\label{47}
dW^{(2\gamma)}_{3p,2p}=\frac{8\omega^3(\omega_0-\omega)^3}{3^5 25\pi}\alpha^2
\left(25I_0^2(\nu) + 76I_2^2(\nu) + 
   25I_0^2(\nu') + 160I_0(\nu') I_2(\nu') + 76I_2^2(\nu') + 
\right.   
   \\
   \nonumber
\left.
  + 10 I_0(\nu)(16I_2(\nu) + 15I_0(\nu') + 6I_2(\nu')) + 12 I_2(\nu)(5I_0(\nu') + 21I_2(\nu'))\right)d\omega\;,
\end{eqnarray}
The integrals $ I_0 $, $ I_2 $ in Eq.(\ref{47}) are defined as
\begin{eqnarray}
\label{48}
I_{0}(\nu)=  \int\limits_0^{\infty}\int\limits_0^{\infty}dr_1dr_2r^3r^3R_{21}(r_1)g_0(\nu;r_1,r_2)R_{31}(r_2)\,\;,
\\
\nonumber
I_2(\nu)= \int\limits_0^{\infty}\int\limits_0^{\infty}dr_1dr_2r^3r^3R_{21}(r_1)g_2(\nu;r_1,r_2)R_{31}(r_2)\,\;,
\end{eqnarray}
$\nu=Z/\sqrt{E_{3p}-\omega}$, $\nu'=Z/\sqrt{E_{3p}-\omega'}$, $\omega'=\omega_0-\omega$
and the radial part $g_l(\nu; r,r')$ of the Coulomb Green function is
\begin{eqnarray}
\label{49}
g_l(\nu; r,r')=\frac{4Z}{\nu}\left(\frac{4}{\nu^2}rr'\right)^l
\exp\left(-\frac{r+r'}{\nu}\right)\sum\limits_{n=0}^{\infty}\frac{n!L^{2l+1}_n\left(\frac{2r}{\nu}\right)L^{2l+1}_n\left(\frac{2r'}{\nu}\right)}{(2l+1+n)!(n+l+1-\nu)}\;.
\end{eqnarray}
In Eq. (\ref{49}) $ L^{2l+1}_{n} $ are the Laguerre polynomials
The radiation "escape" via the 3-photon transition $3p\rightarrow 1s+3\gamma(E1)$ should be defined as
\begin{eqnarray}
\label{50}
Y^{(2)3\gamma}_{3p,1s}=1-X^{(2)3\gamma}_{3p,1s}\;,
\end{eqnarray}
\begin{eqnarray}
\label{51}
X^{(2)3\gamma}_{3p,1s}=\frac{1}{2}\int\limits_0^{\omega_0}I_{3p,2p}(\omega)\left[L_{3p,2s}(\omega)+L_{3s,2p}(\omega)+L_{3d,2p}(\omega)\right]d\omega+\frac{W_{3p,2s}^{(1\gamma)}}{\Gamma_{3p}}X_{2s,1s}^{(2)}\;.
\end{eqnarray}

In Eq. (\ref{51}) the decay rate $W_{3p,2s}^{(1\gamma)}\equiv \Gamma_{3p,2s}^{(1\gamma)}= 2.24603\cdot 10^7$ $s^{-1}$, the decay rate $\Gamma_{3p,1s}^{(1\gamma)}=1.67342\cdot 10^8$ $s^{-1}$ and according to Eq. (\ref{42}) $\Gamma_{3p}=1.89803 \cdot 10^8$ $s^{-1}$. Note that the difference between $W^{(3\gamma)}_{3p,1s}$ 
and $\Gamma_{3p}$ becomes much more significant than the difference between $W^{(2\gamma)}_{3d,1s}$, $W^{(2\gamma)}_{4s,1s}$, $W^{(2\gamma)}_{4d,1s}$ and $\Gamma_{3d}$, $\Gamma_{4s}$, $\Gamma_{4d}$ respectively. The direct one-photon decays of the levels $3d$, $4s$, $4d$ are very much faster than the corresponding cascades (see Table 1). This happens because these cascades contain a "pure two-photon" link, which transition rate is as slow as the $2s-1s$ transition rate. Another situation occurs for the the 3-photon decays $4p\rightarrow1s+3\gamma(E1)$, $4f\rightarrow1s+3\gamma(E1)$. The formulas, similar to Eq.(\ref{41}), for these transition rates look like
\begin{eqnarray}
\label{77a}
W^{(3\gamma)}_{4p,1s}=W^{(2\gamma)}_{4p,2p}+\frac{W^{(1\gamma)}_{3p,1s}}{\Gamma_{3p}}W^{(2\gamma)}_{4p,3p}+\frac{W^{(1\gamma)}_{4p,3d}}{\Gamma_{4p}}W^{(2\gamma)}_{3d,1s}+\frac{W^{(1\gamma)}_{4p,3d}}{\Gamma_{4p}}W^{(2\gamma)}_{3d,1s}+\frac{W^{(1\gamma)}_{4p,3s}}{\Gamma_{4p}}W^{(2\gamma)}_{3s,1s}+\frac{W^{(1\gamma)}_{4p,2s}}{\Gamma_{4p}}W^{(2\gamma)}_{2s,1s}\;,
\end{eqnarray}
\begin{eqnarray}
\label{77b}
W^{(3\gamma)}_{4f,1s}=W^{(2\gamma)}_{4f,2p}+W^{(2\gamma)}_{3d,1s}+\frac{W^{(1\gamma)}_{3p,1s}}{\Gamma_{3p}}W^{(2\gamma)}_{4f,3p}\;.
\end{eqnarray}
The two-photon links of the cascades $4p\rightarrow2p+2\gamma(E1)\rightarrow1s+3\gamma(E1)$, $4f\rightarrow2p+2\gamma(E1)\rightarrow1s+3\gamma(E1)$ are similar to $3s\rightarrow1s+2\gamma(E1)$ transition, i.e. they represent themselves the cascade transitions: $4p\rightarrow3d\rightarrow2p$ and $4f\rightarrow3d\rightarrow2p$ transitions, respectively. Accordingly, these links have to be evaluated in the same way as $3s\rightarrow1s+2\gamma(E1)$ transition and are comparable by the magnitude. With it again the total decay rates $W^{(2\gamma)}_{4p,2p}$, $W^{(2\gamma)}_{4f,2p}$ differ from the total widths $\Gamma_{4p}=\Gamma^{(1\gamma)}_{4p,1s}+\Gamma^{(1\gamma)}_{4p,2s}+\Gamma^{(1\gamma)}_{4p,3s}+\Gamma^{(1\gamma)}_{4p,3d}$ 
and $\Gamma_{4f}=\Gamma^{(1\gamma)}_{4f,3d}$, respectively. The two other decay channels of $4p$, $4f$ levels, i.e.  $4p\rightarrow3p+2\gamma(E1)\rightarrow1s+3\gamma(E1)$ and $4f\rightarrow3p+2\gamma(E1)\rightarrow1s+3\gamma(E1)$ contain "pure two-photon" cascade links and, therefore, have very low transition rates compared to $\Gamma_{4p}$, $\Gamma_{4f}$. The results for the $X_{4p,1s}^{(2)3\gamma}$, $Y_{4p,1s}^{(2)3\gamma}$ and $X_{4f,1s}^{(2)3\gamma}$, $Y_{4f,1s}^{(2)3\gamma}$ are also given in Table 1.

\subsection{Radiation "escape" in the four-photon transitions}

Now we turn to the 4-photon processes and consider first $4s\rightarrow 1s+4\gamma(E1)$ process. This process can occur parallel to the 2-photon process $4s\rightarrow 1s+2\gamma(E1)$ process, described in Section III and does not interfere with the latter due to the different number of the  photons in the final state. As in the case of 3-photon decays we neglect the contribution of the "pure" 4-photon decays and consider only the cascade contributions with 2-photon links, namely $4s\rightarrow 3p+\gamma(E1)\rightarrow 2s+2\gamma(E1)\rightarrow 1s+4\gamma(E1)$, $4s\rightarrow 3s+2\gamma(E1)\rightarrow 2p+3\gamma(E1)\rightarrow 1s+4\gamma(E1)$ and $4s\rightarrow 3p+\gamma(E1)\rightarrow 2p+3\gamma(E1)\rightarrow 1s+4\gamma(E1)$. Then the expression similar to Eq. (\ref{41}) arises \cite{SL}, \cite{SLCan}:
\begin{eqnarray}
\label{52}
W^{(4\gamma)}_{4s,1s}=W^{(2\gamma)}_{4s,3s}+\frac{W^{(1\gamma)}_{4s,3p}}{\Gamma_{4s}}W^{(2\gamma)}_{3p,2p}+\frac{W^{(1\gamma)}_{4s,3p}}{\Gamma_{4s}}\frac{W^{(1\gamma)}_{3p,2s}}{\Gamma_{3p}}W^{(2\gamma)}_{2s,1s}\;.
\end{eqnarray}
In Eq. (\ref{52}) $W_{4s,3p}^{(1\gamma)}\equiv \Gamma_{4s,3p}^{(1\gamma)}$ and the value of $\Gamma_{4s}$ is equal to
\begin{eqnarray}
\label{53}
\Gamma_{4s}=\Gamma_{4s,3p}^{(1\gamma)}+\Gamma_{4s,2p}^{(1\gamma)}\;.
\end{eqnarray}
Similarly, $W_{3p, 2s}^{(1\gamma)}\equiv \Gamma_{3p,2s}^{(1\gamma)}$ and $\Gamma_{3p}$ is defined by Eq. (\ref{42}).

The two-photon differential transition rate (frequency distribution) $dW^{(2\gamma)}_{4s,3s}$ can be obtained from Eqs (\ref{17}), (\ref{18}) with the replacement $2s\rightarrow 4s$, $1s\rightarrow 3s$ and the value of the decay rate magnitude can be found in \cite{SLCan}. Then the radiation "escape" via the 4-photon transition $4s\rightarrow 1s+4\gamma(E1)$ is
\begin{eqnarray}
\label{54}
Y_{4s,1s}^{(2)4\gamma}=1-X_{4s,1s}^{(2)4\gamma}\;,
\end{eqnarray}
\begin{eqnarray}
\label{55}
X_{4s,1s}^{(2)4\gamma}=\frac{1}{2}\int\limits_0^{\omega_0}I_{4s,3s}(\omega)\left[L_{4s,3p}(\omega)+L_{4p,3s}(\omega)+L_{4f,3d}(\omega)+L_{4d,3p}(\omega)+L_{4p,3d}(\omega)\right]d\omega+
\\
\nonumber
+\frac{1}{2}\frac{W^{(1\gamma)}_{4s,3p}}{\Gamma_{4s}}\int\limits_0^{\omega_0}I_{3p,2p}(\omega)\left[L_{3p,2s}(\omega)+L_{3s,2p}(\omega)+L_{3d,2p}(\omega)\right]d\omega+\frac{W^{(1\gamma)}_{4s,3p}}{\Gamma_{4s}}\frac{W^{(1\gamma)}_{3p,2s}}{\Gamma_{3p}}X_{2s,1s}^{(2)2\gamma}\;,
\end{eqnarray}
where $\omega_0=E_{4s}-E_{1s}$. Inserting all the necessary frequency distributions for the one-photon transitions in Eq. (\ref{55}), we obtain the result, given in Table 1.

The results of similar derivations performed for the 4-photon $4d\rightarrow 1s+4\gamma(E1)$ transition see also in Table 1. Note that for these $4$-photon decays as for the 3-photon decay $3p\rightarrow1s+3\gamma(E1)$ the total decay rates are much smaller than the total level widths $\Gamma_{4s}$, $\Gamma_{4d}$. With the calculations performed in Sections III, IV all the contributions to the radiation "escape" from $nl(n\leq 4)$ levels in the "two-photon approximation" are exhausted.

\section{The role of the excited levels in the radiation "escape" in the epoch of the cosmological recombination}

In this Section we will estimate the relative importance of the multiphoton (two-, three-, four-photon) decays of the $ns, np  (2<n\leq 4)$, $nd$, $nf(n\leq 4)$ levels in the radiation "escape" in the epoch of the cosmological recombination. We assume, as usually that the thermodynamical equilibrium existed and the electron temperature $T_e$ was approximately equal to the photon temperature $T_{\gamma}$ \cite{Seager}. For defining $T_{\gamma}$ we employ the formula \cite{Seager}
\begin{eqnarray}
\label{56}
T_{\gamma}=T_0(1+z)\;,
\end{eqnarray}
where $T_0=2.725K$ is the recent CMB temperature and $z$ is the redshift, which for the estimates can be taken as $z\approx 1000$ for the cosmological recombination epoch.

Our aim is to compare the role of the excited states in the radiation "escape" with the well known role of the $2s$ level \cite{Zeld}, \cite{Peebles}. For this comparison any characteristics of the radiation "escape" can be employed, which we will not specify here. Denoting this characteristics as $R_{2s}$ we can suggest the following formula for the estimate of the relative role of the excited states
\begin{eqnarray}
\label{57}
R=R_{2s}\left[1+\left(Y_{3s,1s}^{(2)2\gamma}+\frac{g_d}{g_s}Y_{3d,1s}^{(2)2\gamma}\right)e^{-\frac{E_{3s}-E_{2s}}{kT_e}}+
\right.
\\
\nonumber
\left.
+\left(Y_{4s,1s}^{(2)2\gamma}+\frac{g_d}{g_s}Y_{4d,1s}^{(2)2\gamma}+\frac{g_p}{g_s}Y_{4p,1s}^{(2)3\gamma}+\frac{g_f}{g_s}Y_{4f,1s}^{(2)3\gamma}\right)e^{-\frac{E_{4s}-E_{2s}}{kT_e}}\right]\;.
\end{eqnarray}
In Eq. (\ref{57}) the degeneracy numbers for $nl$ states are: $g_s=2$, $g_p=6$, $g_d=10$, $g_f=14$. The "escape" probabilities $Y_{nl,1s}$ are compiled in the Table 1. For the rough estimate we can set all the $Y_{nl,1s}^{(2)s\gamma}$ values equal to 1 (see Table 1). For $T_e=2725 K$ the first exponential factor in the right-hand side of Eq. (\ref{57}) equals to
\begin{eqnarray}
\label{58}
e^{-\frac{E_{3s}-E_{2s}}{kT_e}}\approx 0.00032105
\end{eqnarray}
and the second exponential factor equals to
\begin{eqnarray}
\label{59}
e^{-\frac{E_{4s}-E_{2s}}{kT_e}}\approx 0.000019225\;.
\end{eqnarray}
Hence,
\begin{eqnarray}
\label{60}
R=R_{2s}\left[1+0.0018501+0.0002941\right]=R_{2s}\left[1+0.0021442\right]\;.
\end{eqnarray}
The numbers in of Eq. (\ref{58}) mean that the radiation "escape" from all the excited levels with $n=3$ can contribute at the level $0.18\%$ and the radiation "escape" from all the levels with $n=4$ can contribute about $0.03\%$. Both these numbers may be essential on the recent level of accuracy of the astrophysical observations. It also evident that the levels with $n>4$ can hardly give sizeable contribution. From this formula we excluded $3$-photon decay of $3p$ level and $4$-photon decays of $\Gamma_{4s}$, $\Gamma_{4d}$ levels which were considered in subsections A, B of Section IV. The branching ratios for these decays, according to the Table 1, are too small. To contribute essentially to the "escape" probability the cascade transition has to contain two-photon link which is not "pure" two-photon. The only $3$-photon decays which satisfy  this condition are the decays of $\Gamma_{4p}$, $\Gamma_{4f}$ levels. In other words essential contribution comes from the $3$-photon cascade processes. The $4$-photon decays, compatible with this condition can arise only for the $nl$ levels with $n\geqslant5$, for example $5s\rightarrow1s+4\gamma(E1)$.
\section{Conclusion}

In this paper we have presented a simple model, based on QED, to estimate the relative role of the excited level few photon decays in the radiation "escape" in the cosmological recombination epoch. These estimates cannot replace the accurate solution of the astrophysical balance equations but can give a hint which processes (decays) should be included in the these equations. In particular, it appears that the 3-photon cascade decays of the $nl(n\leq 4)$ levels can give a contribution comparable with the widely discussed contribution of the two-photon decays. Our studies are based on the "two-photon approximation" when we take into account the cascades which apart from the one-photon links have also one two-photon link. This approximation seems to be well justified due to the relative smallness of the "pure" 3-photon, 4-photon etc. processes and due to the smallness of the processes with several two-photon cascade links.

In our investigations we also used an idea of the photon reemission probability first introduced in QED by F. Low \cite{Low}. According to this idea, if an atom emits a photon which frequency is distributed as a result of the preceding absorption, the total probability of the photon emission can be smaller then 1. The deviation of this probability from the unity reflects the fact that the incoming photon was not necessarily absorbed. This deviation we associate with the radiation "escape". We assume also, that the use of one re-emission ($Y_{nl,1s}^{(2)k\gamma}$ value) is enough to characterize the importance of a certain decay channel for the radiation "escape".

The total contribution of the excited $ns(n>2)$, $nl(l-1,2,3,4, n\leq 4)$ levels to the radiation "escape" compared to the contribution of $2s\rightarrow 1s+2\gamma(E1)$ process according to our estimates reaches the value $0.21\%$ which is not negligible in view of the growing accuracy of the recent astrophysical observations. The smallness of this contribution is due exclusively to the relatively low equilibrium temperature during the epoch of the cosmological recombination.

\newpage
\begin{table}[hP]
\caption{
Contributions of the multiphoton cascade processes, having one two-photon link, to the radiation "escape". Here $k$ is the number of photons; $nl$ - initial state of an atom; $W^{(k\gamma)}_{nl,n'l'}$ - total transition rate for the cascade transition;$W^{(k\gamma)*}_{nl,n'l'}$ - transition rate via "pure two-photon links", $\Gamma_{nl}$ - total width of the upper level in the two-photon link; $X^{(k\gamma)}_{nl,n'l'}$ - reemission probability for the photons from the two-photon link of the $k$-photon cascade; $Y^{(k\gamma)}_{nl,n'l'}$ - "escape" probability for the photons from the two-photon link of the $k$-photon cascade.}
\label{tab:1}
\begin{tabular}{| l | c || c | c | c | c | c | r |}

\hline\hline
$$&$k$ 
&$nl$ &$W^{(k\gamma)}_{nl,1s}, s^{-1}$&$W^{(k\gamma)*}_{nl,1s}, s^{-1}$&$\Gamma_{nl}, s^{-1}$& $X^{(2)k\gamma}_{nl1s}$\qquad& $Y^{(2)k\gamma}_{nl1s}$ \qquad\\
\hline\hline
$1$ & $2$ 
 &  $2s$ &$$&$8.22935$&$8.22935$&$6.39353\times10^{-22}$\qquad&$ 1.00000 $\qquad\\
$2$ & $ $ &  $3s$ &$0.06317\times10^{8}$&$ $&$0.06317\times10^{8}$&$0.00497$\qquad&$ 0.99504 $\qquad\\
$3$ & $ $ &  $3d$ &$0.64686\times10^{8}$&$ $&$0.64686\times10^{8}$&$0.04652$\qquad&$ 0.95349 $\qquad\\
$4$ & $ $ &  $4s$ &$0.04171\times10^{8}$&$ $&$0.04416\times10^{8}$&$0.00431$\qquad&$ 0.99569 $\qquad\\
$5$ & $ $ &  $4d$ &$0.26013\times10^{8}$&$ $&$0.27677\times10^{8}$&$0.02118$\qquad&$ 0.97882 $\qquad\\
\hline
$6$ & $3$
& $3p$ &$ $&$1.01909$&$1.89803\times10^{8}$&$ 2.33432\times10^{-22} $ \qquad&$ 1.00000 $\qquad  \\
$7$ & $ $ & $4p$&$0.003929\times10^{8}$&$ $&$0.81311\times10^{8}$&$0.011005$\qquad&$ 0.98995$\qquad\\
$8$ & $ $ & $4f$ &$0.784812\times10^{8}$&$ $&$0.13795\times10^{8}$&$0.08435$\qquad&$0.95348$\qquad\\
\hline
$9$ & $4$
& $4s$ &$ $&$0.61571$&$0.04416\times10^{8}$& $ 2.47954\times10^{-22} $\qquad&$ 1.00000 $\qquad  \\
$10$ & $ $ & $4d$&$ $&$0.41132$&$0.27678\times10^{8}$&$ 1.75724\times10^{-22} $\qquad&$ 1.00000 $\qquad\\
\hline \hline
\end{tabular}
\end{table}

\begin{center}
Acknowledgments
\end{center}
The authors are indebted to V. K. Dubrovich for helpful consultations. This work was supported by RFBR grant Nr. 11-02-00168-a and by Goskontrakt $\Pi$1334. T. Z. acknowledges support by the Non-profit Foundation "Dynasty" (Moscow).


\setcounter{equation}{0}
\renewcommand{\theequation}%
{A.\arabic{equation}}

\section*{Appendix A: Resonant scattering of photons on an atomic electron and the line profile for the emission process}
We consider first a $ n $-photon elastic scattering process, depicted in the Feynman graph Fig.1.
The corresponding S-matrix element can be written as
\begin{eqnarray}
\label{a1}
\hat{S}^{(2n)}  = (-ie)^{2n}\int d^4x_1d^4x_2...d^4x_{2n-1}d^4x_{2n}\bar{\psi}_{a_0}(x_1)\gamma_{\mu_1}A^{*(k_1e_1)}_{\mu_1}(x_1)S(x_1x_2)\gamma_{\mu_2}A^{*(k_2e_2)}_{\mu_2}(x_2)S(x_2x_3)\dotsc\nonumber \\\gamma_{\mu_n}A^{*(k_ne_n)}_{\mu_n}(x_n)S(x_nx_{n+1})\gamma_{\mu_{n+1}}A^{(k_{n+1}e_{n+1})}_{\mu_{n+1}}(x_{n+1})S(x_{n+1}x_{n+2})
\dotsc\gamma_{\mu_{2n-1}}A^{k_{2n-1}e_{2n-1}}_{\mu_{2n-1}}(x_{2n-1})\nonumber \\S(x_{2n-1}x_{2n})\gamma_{\mu_{2n}}A^{(k_{2n} e_{2n})}_{\mu_{2n}}(x_{2n})\psi_{a_0}(x_{2n})\; ,
\end{eqnarray}
where $\hat{S}^{(2n)}$ is the $S$-matrix of $ 2n $-order, $e$ is the charge of electron, $\psi_a(x) = \psi_a(\vec{r})e^{-i E_a t}$ is the solution of the Dirac equation for the atomic electron, $E_a$ is the Dirac energy. $\bar{\psi}_{a}(x)= \psi_{a}^\dagger \gamma_0$ is the Dirac conjugated wave function with $\psi_{a}^{\dagger}$ being it's Hermitian conjugate, $\gamma_{\mu} = (\gamma_0, \vec\gamma)$  are the Dirac matrices. Wave function of photon $A_{\mu}(x)$ looks like:
\begin{eqnarray}
\label{a2}
A^{(\vec e,\,\vec k)}_{\mu}(x) = \sqrt{\frac{2\pi}{\omega}}\,e^{(\lambda)}_{\mu}e^{i(\vec{k}\vec{r}-\omega t)}
 = \sqrt{\frac{2\pi}{\omega}}\,e^{(\lambda)}_{\mu}e^{-i\omega t}\,A^{(\vec e,\,\vec k)}_{\mu}(\vec r\,)
\; ,
\end{eqnarray}
where $e^{(\lambda)}_{\mu}$ is the photon polarization four-vector, $k=(\vec{k},\omega)$ - is the photon momentum four-vector ($\vec{k}$ is the wave vector, $\omega=|\vec{k}|$ is the photon frequency), $x\equiv(\vec{r},t)$ is the coordinates four-vector ($\vec{r}, t$ are the space and time coordinates). Function (\ref{a2}) corresponds to the absorbed photon and the function $A^*_{\mu}(x)$ corresponds to the emitted one.
$S(x_1x_2)$ is the Feynman propagator for atomic electron. In the Furry picture, the eigenmode decomposition for this propagator reads:
\begin{eqnarray}
\label{a3}
S(x_1x_2)=\frac{i}{2\pi}\int\limits_{-\infty}^{\infty}d\omega_1e^{i\omega_1(t_1-t_2)}\sum\limits_s\frac{\psi_s(\vec{r}_1)\bar{\psi}_s(\vec{r}_2)}{\omega_1-E_s(1-i0)}\; ,
\end{eqnarray}
where summation in (\ref{a3}) extends over the entire Dirac spectrum of the electron states $s$ in the field of nucleus.
Integration over frequency and time variables in (\ref{a1}) leads to
\begin{eqnarray}
\label{a4}
\hat{S}^{(2n)}=(-2\pi i)(e)^{2n}\delta\left(\sum_{i=1}^n\omega_i -\sum_{i=n+1}^{2n}\omega_i \right)\sum_{\substack{s_1,s_2,\dotsc s_{n},\\s_{n+1},\dotsc ,s_{2n-1}}}\frac{(\vec{\alpha}\vec{A}^{*(e_1,\,k_1)})_{a_0s_1}(\vec{\alpha}\vec{A}^{*(e_2,\,k_2)})_{s_1s_2}\dotsc(\vec{\alpha}\vec{A}^{*(e_n,\,k_n)})_{s_{n}s_{n+1}}}{(E_{a_0}-E_{s_1}+\omega_1)(E_{a_0}-E_{s_2}+\omega_1+\omega_2)\dotsc}\nonumber\\\times\frac{(\vec{\alpha}\vec{A}^{(e_{n+1},\,k_{n+1})})_{s_{n+1}s_{n+2}}\dotsc(\vec{\alpha}\vec{A}^{(e_{2n-1},\,k_{2n-1})})_{s_{2n-1}s_{2n}}(\vec{\alpha}\vec{A}^{(e_{2n},\,k_{2n})})_{s_{2n}a_0}}{(E_{a_0}-E_{s_n}+\sum_{i=1}^n\omega_i)(E_{a_0}-E_{s_{n+1}}+\sum_{i=n+2}^{2n}\omega_i)\dotsc(E_{a_0}-E_{2n-1}+\omega_{2n})}\;.
\end{eqnarray}

The amplitude of the process of elastic photon scattering is related to the $ S $-matrix element via
\begin{eqnarray}
\label{a5}
\hat{S}^{(2n)}=(-2\pi i)\delta\left(\sum_{i=1}^n\omega_i -\sum_{i=n+1}^{2n}\omega_i \right)U^{sc.(2n)}\;.
\end{eqnarray}
The energy conservation in this process is implemented by the condition
\begin{eqnarray}
\label{a6}
\sum_{i=1}^n\omega_i=\sum_{i=n+1}^{2n}\omega_i 
\end{eqnarray}
and the resonance frequencies are given by
\begin{eqnarray}
\label{a7}
\omega_{2n}=\omega_1=E_{a_1}-E_{a_0}\nonumber\\
\omega_{2n-1}=\omega_2=E_{a_2}-E_{a_1}\nonumber\\
........................\nonumber\\
\omega_{n+1}=\omega_n=E_{a_n}-E_{a_{n-1}}\nonumber\\
(E_{a_n}\equiv E_a)\;.
\end{eqnarray}

Accordingly, we will obtain for the scattering amplitude the expression
\begin{eqnarray}
\label{a10}
U^{sc. (2n)}=\sum_{\substack{s_1,s_2,\dotsc s_{n},\\s_{n+1},\dotsc ,s_{2n-1}}}\frac{(U^*_{\omega_1})_{a_0s_1}(U^*_{\omega_2})_{s_1s_2}\dotsc(U^*_{\omega_n})_{s_{n-1}s_{n}}}{(E_{a_0}-E_{s_1}+\omega_1)(E_{a_0}-E_{s_2}+\omega_1+\omega_2)\dotsc(E_{a_0}-E_{s_n}+\sum_{i=1}^n\omega_i)}\nonumber\\\times\frac{(U_{\omega_{n+1}})_{s_{n}s_{n+1}}\dotsc(U_{\omega_{2n-1}})_{s_{2n-2}s_{2n-1}}(U_{\omega_{2n}})_{s_{2n-1}a_0}}{(E_{a_0}-E_{s_{n+1}}+\sum_{i=1}^{n}\omega_i-\omega_{n+1})\dotsc(E_{a_0}-E_{s_{2n-1}}+\sum_{i=1}^{n}\omega_i-\sum_{i=n+1}^{2n-1}\omega_i)}\;,
\end{eqnarray}
where we abbreviate the transition matrix element as
\begin{eqnarray}
\label{a11}
(U_{\omega})_{ab}\equiv(\vec{\alpha}A^{(e,\,k)})_{ab}\;.
\end{eqnarray}
In the resonance case we have to retain the terms $s_1=a_1$, $s_2=a_2$, $\dotsc $ , $ s_n=a $,  $\dotsc $, $ s_{2n-2}=a_2$, $ s_{2n-1}=a_1 $ in Eq. (\ref{a10}), which yields 
\begin{eqnarray}
\label{a12}
U^{sc.(2n),res}= \frac{(U^*_{\omega_1})_{a_0a_1}(U^*_{\omega_2})_{a_1a_2}\dotsc(U^*_{\omega_n})_{a_{n-1}a}}{(E_{a_0}-E_{a_1}+\omega_1)(E_{a_0}-E_{a_2}+\omega_1+\omega_2)\dotsc(E_{a_0}-E_{a_n}+\sum_{i=1}^n\omega_i)}\nonumber\\\times\frac{(U_{\omega_{n}})_{aa_{n-1}}\dotsc(U_{\omega_{2}})_{a_2a_1}(U_{\omega_{1}})_{a_1a_0}}{(E_{a_0}-E_{a_{n-1}}+\sum_{i=1}^{n-1}\omega_i)(E_{a_0}-E_{a_{n-2}}+\sum_{i=1}^{n-2}\omega_i)\dotsc(E_{a_0}-E_{a_1}+\omega_{1})}\;,
\end{eqnarray}
In order to describe the line profile for the multiphoton emission, we have to consider first the amplitude for the multiphoton emission in the resonance approximation, which can be defined as
\begin{eqnarray}
\label{a13}
U^{em}=\frac{(U^*_{\omega_1})_{a_0a_1}(U^*_{\omega_2})_{a_1a_2}\dotsc(U^*_{\omega_n})_{a_{n-1}a}}{(E_{a_0}-E_{a_1}+\omega_1)(E_{a_0}-E_{a_2}+\omega_1+\omega_2)\dotsc(E_{a_0}-E_{a_n}+\sum_{i=1}^n\omega_i)}\;.
\end{eqnarray}
The resonance approximation for this $ n $-photon emission process assumes actually the existence of the cascade transition $a\rightarrow a_{n-1}\rightarrow a_{n-2}\rightarrow\dotsc\rightarrow a_1\rightarrow a_0$. The problem of cascades will be investigated below. An expression, similar to Eq.(\ref{a13}) can be defined for the corresponding absorption amplitude
\begin{eqnarray}
\label{a14}
U^{ab}=\frac{(U_{\omega_{n}})_{aa_{n-1}}\dotsc(U_{\omega_{2}})_{a_2a_1}(U_{\omega_{1}})_{a_1a_0}}{(E_{a_0}-E_{a_{n}}+\sum_{i=1}^{n}\omega_i)(E_{a_0}-E_{a_{n-1}}+\sum_{i=1}^{n-1}\omega_i)\dotsc(E_{a_0}-E_{a_1}+\omega_{1})}\;.
\end{eqnarray}

The factors in denominators of Eq.(\ref{a13}) generate simple poles at the resonance frequencies. These singularities are removed (from the real axis) by inserting radiative corrections into the central and all upper electron propagators, i.e. by introducing "radiative dressed" propagators. The insertion of the lowest-order radiative corrections (the vacuum polarization, represented by the Uehling potential, and the electron self-energy) in the central propagator (see Fig.2) yields 
\begin{eqnarray}
\label{a15}
U^{sc. (2n+2)}=\sum_{\substack{s_1,s_2,\dotsc s_{n},\\s_{n+1},\dotsc ,s_{2n-1},s_{2n}}}\frac{(U^*_{\omega_1})_{a_0s_1}(U^*_{\omega_2})_{s_1s_2}\dotsc(U^*_{\omega_n})_{s_{n-1}s_{n}}}{(E_{a_0}-E_{s_1}+\omega_1)(E_{a_0}-E_{s_2}+\omega_1+\omega_2)\dotsc(E_{a_0}-E_{s_n}+\sum_{i=1}^n\omega_i)}\nonumber\\\times \left[\frac{(\widehat{\Sigma}(\sum_{i=1}^n\omega_i)_{s_{n}s_{n+1}}+V^U_{s_{n}s_{n+1}})_{s_{n}s_{n+1}}}{E_{a_0}-E_{s_{n+1}}+\sum_{i=1}^n\omega_i}\right]\nonumber\\\times\frac{(U_{\omega_{n+1}})_{s_{n+1}s_{n+2}}\dotsc(U_{\omega_{2n-1}})_{s_{2n-1}s_{2n}}(U_{\omega_{2n}})_{s_{2n}a_0}}{(E_{a_0}-E_{s_{n+2}}+\sum_{i=1}^{n}\omega_i-\omega_{n+1})\dotsc(E_{a_0}-E_{s_{2n}}+\sum_{i=1}^{n}\omega_i-\sum_{i=n+1}^{2n-1}\omega_i)}\;,
\end{eqnarray}

where $V^U$ is the Uehling potential and $ (\widehat{\Sigma}(\xi))_{a_1a_2} $ represents the matrix element of the energy-dependent self-energy operator
\begin{eqnarray}
\label{a16}
(\widehat{\Sigma}(\xi))_{a_1a_2}\equiv e^2\sum\limits_{n}\frac{i}{2\pi}\int d\Omega\frac{(I(|\Omega|))_{a_1nna_2}}{\xi-\Omega-E_n(1-i0)}\;.
\end{eqnarray}
Here the shorthand notation
\begin{eqnarray}
\label{a17}
(I(\Omega))_{a'b'ab}\equiv\sum\limits_{\mu_1\mu_2}\int d^3r_1d^3r_2\overline{\psi}_{a'}(r_1)\overline{\psi}_{b'}(r_2)\gamma^{\mu_1}_1\gamma^{\mu_2}_2I_{\mu_1\mu_2}(\Omega,r_{12})\psi_{a'}(r_1)\psi_{b'}(r_2)\;,
\end{eqnarray}
is used with 
\begin{eqnarray}
\label{a18}
I_{\mu_1\mu_2}(\Omega,r_{12})=g_{\mu_1\mu_2}\frac{1}{r_{12}}e^{i\Omega \,r_{12}}\;,
\end{eqnarray}
$ r_{12}=|r_1-r_2| $ and the metric tensor $ g_{\mu_1\mu_2} $.\\
In the resonance case Eq.(\ref{a15}) reduces to
\begin{eqnarray}
\label{a19}
U^{sc,(2n+2)}=\frac{(U^*_{\omega_1})_{a_0a_1}(U^*_{\omega_2})_{a_1a_2}\dotsc(U^*_{\omega_n})_{a_{n-1}a}}{(E_{a_0}-E_{a_1}+\omega_1)(E_{a_0}-E_{a_2}+\omega_1+\omega_2)\dotsc(E_{a_0}-E_{a}+\sum_{i=1}^n\omega_i)}\nonumber\\\times \left[\frac{(\widehat{\Sigma}(\sum_{i=1}^n\omega_i)_{aa}+V^U_{aa})_{aa}}{E_{a_0}-E_{a}+\sum_{i=1}^n\omega_i}\right]\nonumber\\\times\frac{(U_{\omega_{n}})_{aa_{n-1}}\dotsc(U_{\omega_{2}})_{a_{2}a_{1}}(U_{\omega_{1}})_{a_1a_0}}{(E_{a_0}-E_{a_{n-2}}+\sum_{i=1}^{n-1}\omega_i)\dotsc(E_{a_0}-E_{a_{1}}+\omega_1)}\;.
\end{eqnarray}

Ressumation of an infinite sequence of radiative insertions to all orders of the perturbation theory (geometric progression) leads to the following expression for the emission amplitude
\begin{eqnarray}
\label{a20}
U^{em}=\frac{(U^*_{\omega_1})_{a_0a_1}(U^*_{\omega_2})_{a_1a_2}\dotsc(U^*_{\omega_n})_{a_{n-1}a}}{(E_{a_0}-E_{a_1}+\omega_1)(E_{a_0}-E_{a_2}+\omega_1+\omega_2)\dotsc(E_{a_0}-E_{a}+\sum_{i=1}^n\omega_i-V_a(\sum_{i=1}^n\omega_i))}\;.
\end{eqnarray}
with the (in general complex-valued) energy corrections
\begin{eqnarray}
\label{a21}
V_{a}(\sum_{i=1}^n\omega_i)=E_a+(\widehat{\Sigma}(\sum_{i=1}^n\omega_i))_{aa}+V^U_{aa}\;.
\end{eqnarray}
\\
Expanding the expression for the matrix element of the operator $ \widehat{\Sigma} $ into Taylor series around the resonance energy $E_{a_0}+\sum_{i=1}^n\omega_i=E_{a} $ and retaining only the leading term in the correction Eq.(\ref{a21}) yield 
\begin{eqnarray}
\label{a22}
U^{em}=\frac{(U^*_{\omega_1})_{a_0a_1}(U^*_{\omega_2})_{a_1a_2}\dotsc(U^*_{\omega_n})_{a_{n-1}a}}{(E_{a_0}-E_{a_1}+\omega_1)(E_{a_0}-E_{a_2}+\omega_1+\omega_2)\dotsc(E_{a_0}+\sum_{i=1}^n\omega_i-V_a)}\;, 
\end{eqnarray}
where
\begin{eqnarray}
\label{a23}
V_{a}=E_a+(\widehat{\Sigma}(E_a))_{aa}+V^U_{aa}\;.
\end{eqnarray}
Now let us turn to the insertion in the first upper electron propagator. After performing time and frequency integrations the corresponding $S$-matrix element reads
\begin{eqnarray}
\label{a24}
U^{sc. (2n+2)}=\sum_{\substack{s_1,s_2,\dotsc s_{n},\\s_{n+1},\dotsc ,s_{2n-1},s_{2n}}}\frac{(U^*_{\omega_1})_{a_0s_1}(U^*_{\omega_2})_{s_1s_2}\dotsc(U^*_{\omega_{n-1}})_{s_{n-2}s_{n-1}}}{(E_{a_0}-E_{s_1}+\omega_1)(E_{a_0}-E_{s_2}+\omega_1+\omega_2)\dotsc(E_{a_0}-E_{s_{n-1}}+\sum_{i=1}^{n-1}\omega_i)}\nonumber\\\times \left[\frac{(\widehat{\Sigma}(\sum_{i=1}^{n-1}\omega_i)_{s_{n-1}s_{n}}+V^U_{s_{n-1}s_{n}})_{s_{n-1}s_{n}}}{E_{a_0}-E_{s_{n}}+\sum_{i=1}^{n-1}\omega_i}\right]\nonumber\\\times\frac{(U^*_{\omega_n})_{s_{n}s_{n+1}}(U_{\omega_{n+1}})_{s_{n+1}s_{n+2}}\dotsc(U_{\omega_{2n-1}})_{s_{2n-1}s_{2n}}(U_{\omega_{2n}})_{s_{2n}a_0}}{(E_{a_0}-E_{s_{n+1}}+\sum_{i=1}^{n}\omega_i)(E_{a_0}-E_{s_{n+2}}+\sum_{i=1}^{n}\omega_i-\omega_{n+1})\dotsc(E_{a_0}-E_{s_{2n}}+\sum_{i=1}^{n}\omega_i-\sum_{i=n+1}^{2n-1}\omega_i)}\;.
\end{eqnarray}

The resonant case is characterized by the conditions $ s_{n-1}=s_n=a_{n-1}, s_1=a_1, s_2=a_2,\dotsc, s_{n+1}=a, s_{n+2}=a_{n-1},\dotsc, s_{2n}=a_1$, which imply the corresponding scattering amplitude
\begin{eqnarray}
\label{a25}
U^{res. (2n+2)}=\frac{(U^*_{\omega_1})_{a_0a_1}(U^*_{\omega_2})_{a_1a_2}\dotsc(U^*_{\omega_{n-1}})_{a_{n-2}a_{n-1}}}{(E_{a_0}-E_{a_1}+\omega_1)(E_{a_0}-E_{a_2}+\omega_1+\omega_2)\dotsc(E_{a_0}-E_{a_{n-1}}+\sum_{i=1}^{n-1}\omega_i)}\nonumber\\\times \left[\frac{(\widehat{\Sigma}(\sum_{i=1}^{n-1}\omega_i)_{a_{n-1}a_{n-1}}+V^U_{a_{n-1}a_{n-1}})_{a_{n-1}a_{n-1}}}{E_{a_0}-E_{a_{n-1}}+\sum_{i=1}^{n-1}\omega_i}\right]\nonumber\\\times\frac{(U^*_{\omega_n})_{a_{n-1}a}(U_{\omega_{n}})_{aa_{n-1}}\dotsc(U_{\omega_{2}})_{a_{2}a_{1}}(U_{\omega_{1}})_{a_{1}a_0}}{(E_{a_0}-E_{a}+\sum_{i=1}^{n}\omega_i)(E_{a_0}-E_{a_{n-1}}+\sum_{i=1}^{n-1}\omega_i)\dotsc(E_{a_0}-E_{a_{1}}+\omega_1)}\;,
\end{eqnarray}
We can assume, that all necessary resonant insertions into the central electron propagator in Fig. 2 have been already introduced. Repeating the radiative insertions in the upper electron line in the resonance approximation and performing resummation of the resulting geometrical progression finally we find
\begin{eqnarray}
\label{a26}
U^{em}=\frac{(U^*_{\omega_1})_{a_0a_1}(U^*_{\omega_2})_{a_1a_2}\dotsc(U^*_{\omega_n})_{a_{n-1}a}}{(E_{a_0}-E_{a_1}+\omega_1)\dotsc(E_{a_0}+\sum_{i=1}^{n-1}\omega_i-V_{a_{n-1}}(\sum_{i=1}^{n-1}\omega_i))(E_{a_0}+\sum_{i=1}^n\omega_i-V_a(\sum_{i=1}^n\omega_i))}
\end{eqnarray}
together with
\begin{eqnarray}
\label{a27}
V_{a_{n-1}}(\sum_{i=1}^{n-1}\omega_i)=E_{a_{n-1}}+(\widehat{\Sigma}(\sum_{i=1}^{n-1}\omega_i))_{a_{n-1}a_{n-1}}+V^U_{a_{n-1}a_{n-1}}\;.
\end{eqnarray}
Expanding again the operator $ \widehat{\Sigma} $ in Eq.(\ref{a27}) into a Taylor series close to the point of the resonance, replacing then Eq.(\ref{a27}) by
\begin{eqnarray}
\label{a28}
V_{a_{n-1}}=E_{a_{n-1}}+(\widehat{\Sigma}(E_{a_{n-1}}))_{a_{n-1}a_{n-1}}+V^U_{a_{n-1}a_{n-1}}
\end{eqnarray}
and using also Eq.(\ref{28}), we obtain 
\begin{eqnarray}
\label{a29}
U^{em}=\frac{(U^*_{\omega_1})_{a_0a_1}(U^*_{\omega_2})_{a_1a_2}\dotsc(U^*_{\omega_n})_{a_{n-1}a}}{(E_{a_0}-E_{a_1}+\omega_1)\dotsc(E_{a_0}+\sum_{i=1}^{n-1}\omega_i-V_{a_{n-1}})(E_{a_0}+\sum_{i=1}^n\omega_i-V_a)}\;.
\end{eqnarray}
Inserting radiative corrections into the remaining electron propagators and repeating the procedure described above,  we finally arrive at the following expression for the emission amplitude

\begin{eqnarray}
\label{a30}
U^{em}=\frac{(U^*_{\omega_1})_{a_0a_1}(U^*_{\omega_2})_{a_1a_2}\dotsc(U^*_{\omega_n})_{a_{n-1}a}}{(E_{a_0}+\omega_1-V_{a_1})\dotsc(E_{a_0}+\sum_{i=1}^{n-1}\omega_i-V_{a_{n-1}})(E_{a_0}-E_{a}+\sum_{i=1}^n\omega_i-V_a)}\;, 
\end{eqnarray}
where energies $V_a, V_{a_1}$,..., $V_{a_n}$ are improved by the radiative corrections:
\begin{eqnarray}
\label{a31}
V_{a}=E_{a}+L_{a}-\frac{i}{2}\Gamma_{a}\;,\nonumber \\
V_{a_1}=E_{a_1}+L_{a_1}-\frac{i}{2}\Gamma_{a_1}\;,\nonumber \\
......................................\nonumber \\
V_{a_n}=E_{a_n}+L_{a_n}-\frac{i}{2}\Gamma_{a_n}\;,\nonumber \\
\end{eqnarray}
$L_{a}$, $L_{a_1}$,..., $L_{a_1}$ are the Lamb shifts of the levels $a$, $a_1$,..., $a_n$ and $\Gamma_{a}$, $\Gamma_{a_1}$,..., $\Gamma_{a_n}$ are the corresponding widths.\\
\\
As the next step towards the evaluation of transition probabilities one has to take  the square modulus of $ U^{em} $, to integrate over all emission directions $\vec{\nu}$ and to sum over the polarization $\vec{e}$ of each photon. Defining the partial width $\Gamma_{a_1a_2}$ of the level $a_1$ associated with the transition $a_1\rightarrow a_2$ as
\begin{eqnarray}
\label{a32}
\Gamma_{a_1a_2}=(\omega^{res})^2\sum\limits_{e}\int d^3\nu\vert(U^*_{a_1a_2})\vert^2 \;,
\end{eqnarray}
where $ \omega^{res} $ is the resonant frequency, integrating over all photon directions and summing over polarizations we obtain an expression for the $n$-photon cascade transition probability
\begin{eqnarray}
\label{a33}
dW_{a\rightarrow a_{n-1}\rightarrow\dotsc\rightarrow a_1\rightarrow a_0}=\Bigr(\frac{1}{2\pi}\Bigl)^3\frac{\Gamma_{a_0a_1}}{|E_{a_0}-E_{a_1}+\omega_1-L_{a_1}+\frac{i}{2}\Gamma_{a_1}|^2}\frac{\Gamma_{a_1a_2}}{|E_{a_0}-E_{a_{2}}+\omega_1+\omega_2-L_{a_{2}}+\frac{i}{2}\Gamma_{a_{2}}|^2}\dotsc\nonumber\\\times \frac{\Gamma_{a_{n-2}a_{n-1}}}{|E_{a_0}-E_{a_{n-1}}+\sum_{i=1}^{n-1}\omega_i-L_{a_{n-1}}+\frac{i}{2}\Gamma_{a_{n-1}}|^2}    \frac{\Gamma_{a_{n-1}a}}{|E_{a_0}-E_{a}+\sum_{i=1}^n\omega_i-L_{a}+\frac{i}{2}\Gamma_{a}|^2}\;, 
\end{eqnarray}
where $\Gamma_{a_0a_1}$,..., $ \Gamma_{a_{n-2}a_{n-1}} $ and $\Gamma_{a_{n-1}a}$ are the partial widths, associated with the transition $a_1\rightarrow a_0, $ $ a_2=a_1 $,$\dotsc  $, $ a_{n-1}\rightarrow a_{n-2}$ and $a\rightarrow a_{n-1}$. This is the line profile for the cascades transition, when all photons are assumed to be registered.\\
From the Eq.(\ref{a33}) we can obtain also an expression for single-photon transition $a_1\rightarrow a_o$ in the case when $a_1$,$\dotsc$, $a_{n-1}$, $ a $ states are excited (unstable) states. For this purpose we perform first integration over $\omega_1$ with fixed $\omega_2$, $\omega_3$,..., $ \omega_n $ and after this perform the integration over $\omega_2$ with fixed $\omega_3$, $\omega_4$,..., $ \omega_n $ etc. These integrations can be extended along the entire real axis, since only the pole terms are contributing. We close the contour in the upper half-plane. The poles for the integration over $ \omega_1 $ are located at
\begin{eqnarray}
\label{a34}
\omega^{(1)}_1=E_{a_1}-E_{a_0}+L_{a_1}+\frac{i}{2}\Gamma_{a_1}\nonumber\\
\omega^{(2)}_1=E_{a_{2}}-E_{a_0}-\omega_2+L_{a_{2}}+\frac{i}{2}\Gamma_{a_{2}}\nonumber\\
.............................\nonumber\\
\omega^{(n)}_1=E_{a}-E_{a_0}-\sum_{i=2}^n\omega_i+L_{a}+\frac{i}{2}\Gamma_{a}
\end{eqnarray}
Employing Cauchy's theorem and collecting all residue contributions we repeating this procedure for all remaining frequencies until $ \omega_n $. After some algebraic transformation we finally arrive at the expression

\begin{eqnarray}
\label{a39}
dW_{aa_{n-1}}=\frac{1}{2\pi}\frac{\Gamma_{aa_{n-1}}\Gamma_{a_{n-1}a_{n-2}}..\Gamma_{a_1a_0}(\Gamma_a+\Gamma_{a_{n-1}})}{\Gamma_a\Gamma_{a_{n-1}}..\Gamma_{a_1}}\frac{d\omega}{(\omega-\tilde{\omega}_{aa_{n-1}})^2+\frac{1}{4}(\Gamma_a+\Gamma_{a_{n-1}})^2}\;.
\end{eqnarray}
In Eq.(\ref{a39}) any explicit dependence on the states $ a_1,.., a_{n-2}, a_0 $  has been disappeared. These states enter only implicitly through the definition of the one-photon decay widths. If all states $ a $, $ a_1 $,..., $ a_{n-2} $ decay via single channels i.e. $ \Gamma_{aa_1}=\Gamma_a $, $ \Gamma_{a_1a_2}=\Gamma_{a_1} $ etc., the formula Eq.(\ref{a39}) reduces back to Eq.(3) in the text.


\setcounter{equation}{0}
\renewcommand{\theequation}%
{B.\arabic{equation}}

\section*{Appendix B. Reemission of the photon, emitted by one atom, by another atom: QED derivation}
We start with the derivation of the Lorentz profile for the emission line from the QED description of the photon scattering on the atomic electron \cite{LabShon}, \cite{Low}. \\
The Feynman graph for the one-photon scattering process is depicted in Fig. 1 for $ n=1 $. The resonance approximation is defined by the condition following from Eq. (\ref{a7}):
\begin{eqnarray}
\label{B1}
\omega_1=\omega_2=E_a-E_{a_0}\;.
\end{eqnarray}
A scattering amplitude $ U^{sc.(2)} $ corresponding to Eq. (\ref{a10}) with $ n=1 $ looks like
\begin{eqnarray}
\label{B2}
U^{sc.(2)}=\sum\limits_{s_1}\frac{(U)_{a_0s_1}(U^*)_{s_1a_0}}{E_{a_0}-E_{s_1}+\omega}\;.
\end{eqnarray}
In the resonance approximation 
\begin{eqnarray}
\label{B3}
U^{sc.(2)}=\frac{(U)_{a_0a_1}(U^*)_{a_1a_0}}{E_{a_0}-E_{a_1}+\omega}
\end{eqnarray}
and the emission amplitude is defined as
\begin{eqnarray}
\label{B4}
U^{em.(2)}=\frac{(U^*)_{a_1a_0}}{E_{a_0}-E_{a_1}+\omega}\;.
\end{eqnarray}
Thus the emission amplitude can be obtained from the scattering amplitude in the resonance approximation by omitting absorption matrix element $ (U)_{a_0s_1} $.

A transition probability for one-photon emission process $ a\rightarrow a_0+\gamma $ follows from the Eq (\ref{a39}) and looks like 
\begin{eqnarray}
\label{B5}
dW_{aa_0}=\frac{1}{2\pi}\frac{\Gamma_a d\omega}{(\omega-\tilde{\omega}_{aa_0})+\frac{\Gamma^2_{a}}{4}}\;.
\end{eqnarray}
It is assumed that there are no other decay channels for the transition $ a\rightarrow a_0 $ apart from the one-photon decay $ a\rightarrow a_0+\gamma $ Actually the unique example of such situation is the Lyman-alpha transition $ 2p\rightarrow 1s+\gamma $. The normalization condition Eq. (5) is valid for the Lorentz profile Eq. (\ref{B5}).
\\
Now we will describe the situation when the Lyman-alpha photon, absorbed and emitted by one atom, is then reabsorbed and reemitted by another atom. The corresponding Feynman graph is depicted in Fig. 3. The resonance condition Eq. (B1) is now modified as
\begin{eqnarray}
\label{B6}
\omega'=\omega=E_{a'}-E_{a'_0}=E_{a}-E_{a_0}\;,
\end{eqnarray} 
where quantities with or without "prime"\; index correspond to the electrons in two different atoms.
The S-matrix element, corresponding to the graph Fig. 3 looks like
\begin{eqnarray}
\label{B7}
S^{(4)}=(-ie)^4\int d^4x_1..d^4x_2\overline{\psi}_{a'_0}(x_1)\gamma_{\mu_1}A_{\mu_1}(x_1)S(x_1x_2)\gamma_{\mu_2}\psi_{a'_0}(x_2)D^{t}_{\mu_3\mu_4}\overline{\psi}_{a_0}(x_3)\gamma_{\mu_3}S(x_3x_4)\gamma_{\mu_4}A^*_{\mu_4}(x_4)\psi_{a_0}(x_4)\;.
\end{eqnarray} 
In this matrix element the variables $  x_1$, $x_2 $ correspond to one atom and the variables $  x_3$, $x_4 $ correspond to another one. $D^{t}_{\mu\nu}$ denotes the transverse photon propagator in the Coulomb gauge. We employ this propagator since we want to describe the emission of the real (transverse) photon by one atom and the absorption of this photon by another atom. This propagator can be presented in the form \cite{theSpectra}
\begin{eqnarray}
\label{B8}
D^{t}_{\mu\nu}(x_1,x_2)=\frac{1}{2\pi i}\int_{-\infty}^\infty d\Omega I^{t}_{\mu\nu}(|\Omega|,r_{12})e^{-i\Omega(t_1-t_2)}\;,
\end{eqnarray}
\begin{eqnarray}
\label{bB8}
I^{t}_{\mu_1\mu_2}(\Omega,r_{12})=\left(\frac{\delta_{\mu_1\mu_2}}{r_{12}}e^{i\Omega r_{12}}-\frac{\partial}{\partial^{\mu_1}_1}\frac{\partial}{\partial^{\mu_2}_2}\frac{1}{r_{12}}\frac{e^{i\Omega r_{12}}-1}{\Omega^2}\right)(1-\delta_{\mu_10})(1-\delta_{\mu_20})\;.
\end{eqnarray}
Performing the standard integration over the time and frequency variables and using the relation (\ref{a5}) we arrive as the following expression for the scattering amplitude:
\begin{eqnarray}
\label{B9}
U^{sc. (6)}=e^6\sum\limits_{n,n'}\frac{(\vec{\alpha_1}\vec{A}^{(\vec{e'},\vec{k'})}_{\mu_1})_{a'_0n'}\left(\frac{\vec{\alpha_2}\vec{\alpha_3}}{r_{23}}e^{i(E_{a}-E_{a_0})r_{23}}-(\nabla_{2}\vec{\alpha_2})(\nabla_3\vec{\alpha_3})\frac{1}{r_{23}}\frac{e^{i(E_{a}-E_{a_0})r_{23}}-1}{(E_{a}-E_{a_0})^2}\right)_{a'_0n'na_0}(\vec{\alpha_2}\vec{A}^{*(\vec{e},\vec{k})}_{\mu_4})_{na_0}}{(E_{n'}-E_{a'_0}+\omega')(E_{n}-E_{a_0}+\omega)}\;.
\end{eqnarray}
In the resonance approximation
\begin{eqnarray}
\label{B10}
U^{sc. (6)}=e^6\frac{(\vec{\alpha_1}\vec{A}^{(\vec{e'},\vec{k'})}_{\mu_1})_{a'_0a'}\left(\frac{\vec{\alpha_2}\vec{\alpha_3}}{r_{23}}e^{i(E_{a}-E_{a_0})r_{23}}-(\nabla_{2}\vec{\alpha_2})(\nabla_3\vec{\alpha_3})\frac{1}{r_{23}}\frac{e^{i(E_{a}-E_{a_0})r_{23}}-1}{(E_{a}-E_{a_0})^2}\right)_{a'_0a'aa_0}(\vec{\alpha_2}\vec{A}^{*(\vec{e},\vec{k})}_{\mu_2})_{aa_0}}{(E_{a'}-E_{a'_0}+\omega')(E_{a}-E_{a_0}+\omega)}\;.
\end{eqnarray}

According to Eqs. (\ref{B3})-(\ref{B5}) we obtain the emission amplitude in the resonance approximation by omitting the absorption matrix element in Eq. (\ref{B10})
\begin{eqnarray}
\label{B11}
U^{em}=e^3\frac{\left(\frac{\vec{\alpha_2}\vec{\alpha_3}}{r_{23}}e^{i(E_{a}-E_{a_0})r_{23}}-(\nabla_{2}\vec{\alpha_2})(\nabla_3\vec{\alpha_3})\frac{1}{r_{23}}\frac{e^{i(E_{a}-E_{a_0})r_{23}}-1}{(E_{a}-E_{a_0})^2}\right)_{a'_0a'aa_0}(\vec{\alpha_2}\vec{A}^{*(\vec{e},\vec{k})}_{\mu_2})_{aa_0}}{(E_{a'}-E_{a'_0}+\omega')(E_{a}-E_{a_0}+\omega)}\;.
\end{eqnarray} 
Singularities in the denominators in Eq. (\ref{B10}) should be avoided by the summation of the electron self-energy and vacuum polarization insertions in all the electron lines in Fig. 3. Then, after taking the square modulus of Eq. (\ref{B11}), summation over the polarizations, integration over the angles for emitted photon and finally, after the integration over the frequency $\omega'$ we obtain the following result (setting $ a'_0=a_0=1s $, $ a'=a=2p $)
\begin{eqnarray}
\label{B12}
dW^{(1\gamma)}_{2p-1s}(\omega)=\frac{1}{2\pi}\frac{\Gamma_{2p}d\omega}{\left[(\omega-\omega_{2p1s})^2+\frac{\Gamma_{2p}}{4}\right] ^2} |I_{1s'2p'2p1s}|^2\;,
\end{eqnarray}
where 
\begin{eqnarray}
\label{B13}
I_{1s'2p'2p1s}=e^2\int d\vec{r_2}d\vec{r_3}\psi^*_{2p}(\vec{r_2})\psi^*_{1s}(\vec{r_3})\left(\frac{\vec{\alpha_2}\vec{\alpha_3}}{r_{23}}e^{i(E_{a}-E_{a_0})r_{23}}-(\nabla_{2}\vec{\alpha_2})(\nabla_3\vec{\alpha_3})\frac{1}{r_{23}}\frac{e^{i(E_{a}-E_{a_0})r_{23}}-1}{(E_{a}-E_{a_0})^2}\right)\times\nonumber\\\times\psi_{1s}(\vec{r_2})\psi_{2p}(\vec{r_3})\;.
\end{eqnarray}
In Eq. (\ref{B13}) the one-electron Dirac wave functions $\psi_{2p}$, $ \psi_{1s} $ for the electrons in the two different atoms are present. Thus, the integral (\ref{B13}) depends on the distance between two atoms, i.e. on the density of the atomic gas. It is convenient to fix the origin of the coordinate system at the nucleus of an atom which absorbs the photon.\\
Then we can present $ r_{23} $ in the form
\begin{eqnarray}
\label{bb1}
r_{23}=|\vec{r_2}-\vec{r_3}|=|\vec{R}-\vec{r'}+\vec{r}|\;,
\end{eqnarray}
where $ r' $, $ r $ are the distances between the electrons and the nuclei in the emitting and absorbing atoms, respectively and $ R $ is the distance between the two nuclei. Assuming that $ r' $, $ r\ll R $ we replace the distance $ r_{23} $ by $ R $.\\
We employ equalities
\begin{eqnarray}
\label{bb2}
((\nabla_2\vec{\alpha_2})(\nabla_3\vec{\alpha_3})f(r_{23}))_{A'B'AB}=-([\widehat{\vec{h}_2}[\widehat{\vec{h}_3}f(r_{23})]])_{a'_0a'aa_0}=\nonumber\\
=(E_{a_0}E_{a'}-E_{a_0}E_{a}-E_{a'_0}E_{a'}+E_{a'_0}E_{a})(f(r_{12}))_{a'_0a'aa_0}\;,
\end{eqnarray}
where $ \widehat{h(\vec{r})} $ is the one-electron Dirac Hamiltonian for the bound electron with arbitrary potential $ V(\vec{r}) $
\begin{eqnarray}
\label{bb3}
\widehat{h(\vec{r})}=-i\vec{\alpha}\nabla + \beta m-eV(\vec{r})\;,
\end{eqnarray}
$\vec{\alpha}$, $ \beta $ are the Dirac matrices, $ m $, $ e $ are the mass and the charge of the electron, respectively. The wave functions in the matrix elements are assumed to be the eigenfunctions of the Hamiltonian Eq. (\ref{bb3}) with the eigenvalues $ E_{a_0} $, $ E_{a'_0} $, $ E_a $, $ E_a' $. In case when $ E_{a'_0}=E_{a_0} $, $ E_{a'}=E_{a} $ Eq. (\ref{bb2}) reduces to 
\begin{eqnarray}
\label{bb4}
((\nabla_2\vec{\alpha_2})(\nabla_3\vec{\alpha_3})f(r_{23}))_{a'_0a'aa_0}=(E_{a_0}-E_{a})^2f(r_{23})_{a'_0a'aa_0}\;.
\end{eqnarray}
Now, employing Eq. (\ref{bb4}) with the evident approximation
\begin{eqnarray}
\label{bb5}
\frac{1}{r_{23}}=\frac{1}{R}\;,
\end{eqnarray}
\begin{eqnarray}
\label{bb6}
e^{i(E_{a}-E_{a_0})r_{23}}=e^{i(E_{a}-E_{a_0})R}
\end{eqnarray}
and taking into account the orthogonality of the wave functions $ \psi_{a_0} $, $ \psi_a $, we find that the second term in the brackets in Eq. (\ref{B13}) turns to zero. Then
\begin{eqnarray}
\label{bb7}
I_{1s'2p'2p1s}(R)=\frac{e^2}{R}|(\vec{\alpha}_{2p1s})|^2\;.
\end{eqnarray}
In Eq. (\ref{bb7}) one of the matrix elements $(\vec{\alpha})_{2p1s}$ originates from the emitting atom and another matrix element originates from the absorbing atom. In the nonrelativistic limit \cite{theSpectra}
\begin{equation}
\label{bb8}
|(\vec{\alpha}_{2p1s})|^2\simeq(E_{2p}-E_{1s})^2|(\vec{r})_{2p1s}|^2
\end{equation}
and
\begin{eqnarray}
\label{bb9}
\Gamma_{2p}=\frac{4}{3}e^2(E_{2p}-E_{1s})^3|(\vec{r})_{2p1s}|^2\;.
\end{eqnarray}
Hence
\begin{eqnarray}
\label{bb10}
I_{1s'2p'2p1s}(R)=\frac{3}{4}\frac{\Gamma_{2p}}{(E_{2p}-E_{1s})R}\;.
\end{eqnarray}
We can average the result over the positions of the emitting atoms, surrounding the absorbing atom, assuming the distribution of these atoms spherically symmetrical and introducing the density of the emitting atoms $ \rho(R) $. Then
\begin{eqnarray}
\label{B14}
dW^{1\gamma}_{2p1s}(\omega)=L_{2p1s}(\omega)I_{2p1s}(\omega)\;,
\end{eqnarray}
where
\begin{eqnarray}
\label{B15}
I_{2p1s}(\omega)=\frac{9\pi}{4}\int\limits_0^\infty\frac{\rho(R)R^2dR}{[(E_{2p}-E_{1s})R]^2}\frac{\Gamma^2_{2p}}{(\omega-\tilde{\omega}_{2p1s})^2+\frac{\Gamma^2_{2p}}{4}}\;,
\end{eqnarray}
is the dimensionless function which represents the frequency distribution for the reemitted photon in Eq. (8).  In a simple model, employed in section II B the normalization of the function $ \rho(R) $ was chosen as 
\begin{eqnarray}
\frac{9\pi}{4}\int\limits_0^\infty\frac{\rho(R)R^2dR}{[(E_{2p}-E_{1s})R]^2}=\frac{9\pi}{4(E_{2p}-E_{1s})^2}\int\limits_0^\infty\rho(R)dR=1\;.
\end{eqnarray}

\begin{figure}[H]
\label{image1}
\includegraphics[scale=0.4]{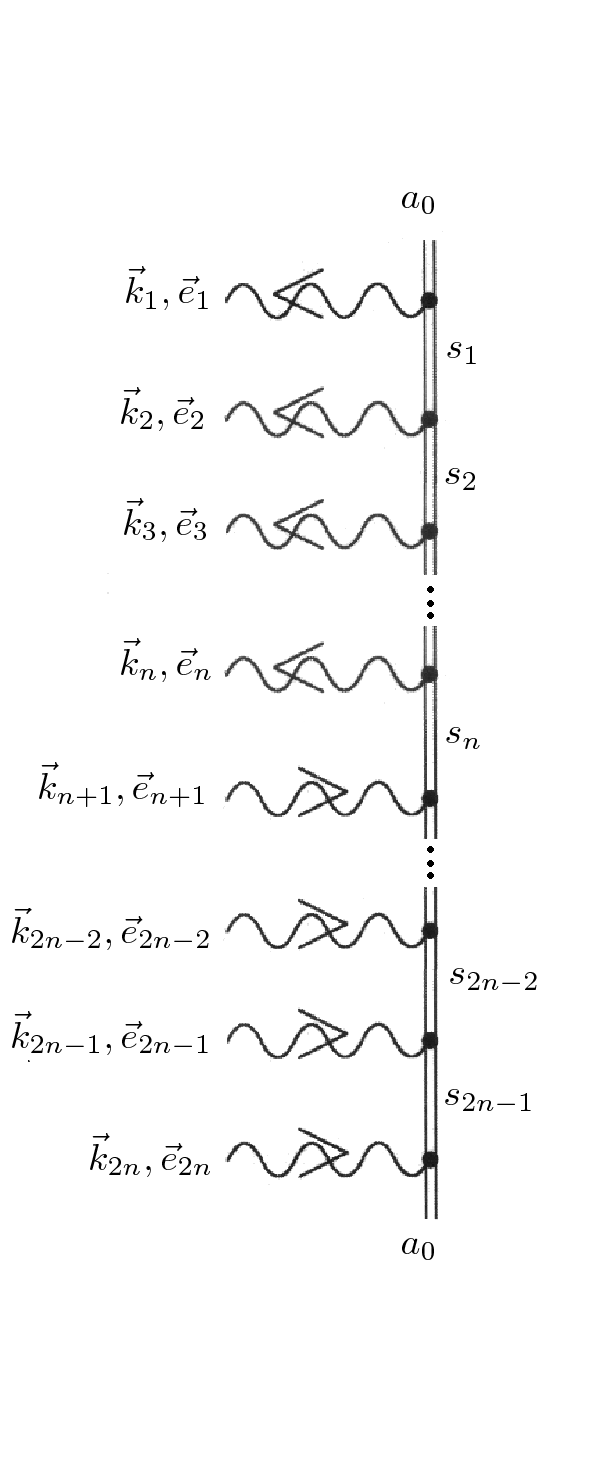}
{\citation 
\\
\\
Fig. 1. Feynman graph describing the elastic scattering of $ n $ photons on an atomic electron in the state $ a_0 $
}
\end{figure}

\begin{figure}[H]
\label{image2}
\includegraphics[scale=0.4]{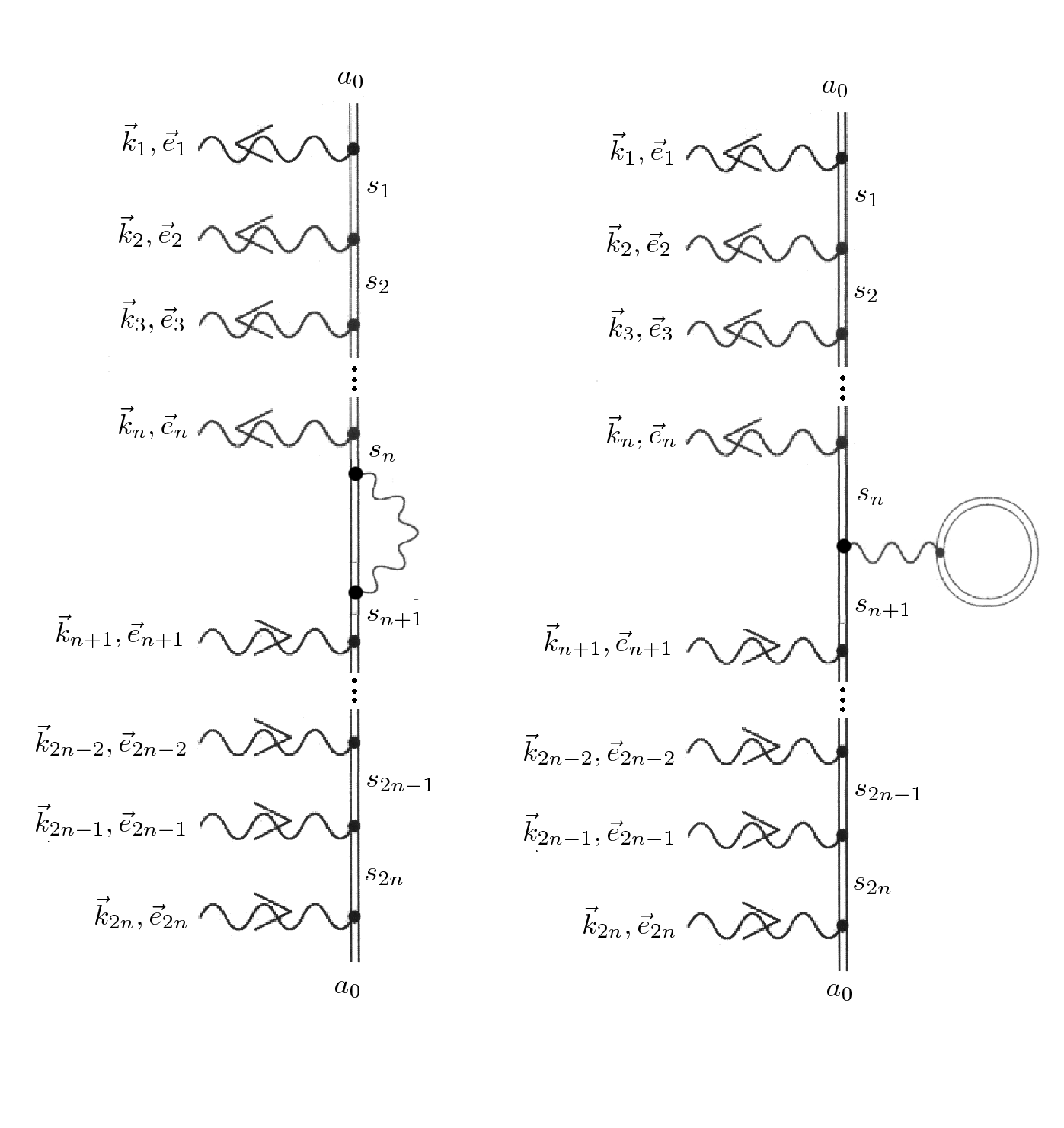}
{\citation 
\\
\\
Fig. 2. Feynman graph corresponding to lowest-order radiative insertions into the central electron propagator in Fig. 1. 
}

\end{figure}

\begin{figure}[H]
\label{image3}
\includegraphics[scale=1]{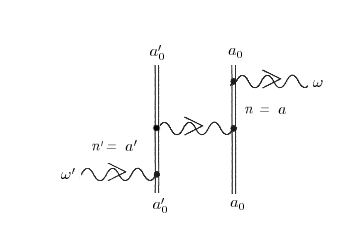}
{\citation 
\\
\\

Fig. 3. The Feynman graph describing the resonance photon rescattering. The two vertical solid lines describe the electrons in two one-electron atoms. The ground and excited states of the one atom are denoted as $a'_0$, $a'$, the states of another atom $a_0$, $a$. The internal wavy line denotes the transverse photon propagation between two atoms. The Coulomb gauge for the photon propagators is assumed.
}
\end{figure}


\newpage
\begin{thebibliography}{99}
\bibitem{Hin} G. Hinshaw, M. R. Nolta, C. L. Bennett et. al., ApJS {\bf 170}, 288 (2007)

\bibitem{Page} L. Page, G. Hinshaw, E. Komatsu et. al., ApJS {\bf 170}, 335 (2007)

\bibitem{Zeld} Ya. B. Zel'dovich, V. G. Kurt and R. A. Syunyaev,
Zh. Eksp. Teor. Fiz. {\bf 55}, 278 (1968) 
[Engl. Transl. Sov. Phys. - JETP Lett. {\bf 28}, 146 (1969)]

\bibitem{Peebles} P. J. E. Peebles, Astrophys. J. {\bf 153}, 1 (1968)

\bibitem{Dubrovich} V. K. Dubrovich and S. I. Grachev, 
Astronomy Letters {\bf 31}, 359 (2006)

\bibitem{Wong} W. Y. Wong and D. Scott, Mon. Not. Roy. Astron. Soc. {\bf 375}, 1441 (2007)

\bibitem{J.Chluba} J. Chluba and R. A. Sunyaev, 
Astronomy and Astrophysics {\bf 480}, 629 (2008)

\bibitem{Hirata} C. M. Hirata, Phys. Rev. D {\bf 78}, 023001 (2008)

\bibitem{Jent1} U. D. Jentschura, J. Phys. A{\bf 40}, F223 (2007)

\bibitem{jas08} U. D. Jentschura and A. Surzhykov, Phys, Rev. A{\bf 77}, 042507 (2008)

\bibitem{LSP} L. Labzowsky, D. Solovyev and G. Plunien, Phys. Rev. A{\bf 80}, 062514 (2009)

\bibitem{Amaro} P. Amaro, J. P. Santos, F. Parente, A. Surzhykov and P. Indelicato,
Phys. Rev. A {\bf 79}, 062504 (2009)

\bibitem{LabShon} L. N. Labzowsky and A. V. Shonin, 
Phys. Rev. A{\bf 69}, 012503 (2004)

\bibitem{ChlubSun} J. Chluba and R. A. Sunyaev, Astronomy and Astrophysics, A{\bf 53}, 512 (2010)

\bibitem{Karsh} S. G. Karshenboim and V. G. Ivanov, Astronomy Letters {\bf 34}, 289 (2009)

\bibitem{WW} V. Weisskopf and E. Wigner, Z. Phys. {\bf 63}, 54 (1930)

\bibitem{Heitler} W. Heitler, The Quantum Theory of Radiation, Oxford 1954

\bibitem{AndrLab} O. Yu. Andreev, L. N. Labzowsky, G. Plunien and D. A. Solovyev,
Phys. Rep. {\bf 455}, 135 (2008) 

\bibitem{Low} F. Low, Phys. Rev. ${\bf 88}$, 53 (1952)


\bibitem{LSPS} L. Labzowsky, D. Solovyev, G. Plunien and G. Soff, Phys. Rev. Lett. {\bf 80}, 143003 (2001)

\bibitem{JM} U. D. Jentschura and P. J. Mohr, Can. J. Phys. {\bf 80}, 633 (2002)

\bibitem{LSS} L. Labzowsky, G. Schedrin, D. Solovyev, E. Chernovskaya, G. Plunien and S. Karshenboim, Phys. Rev. A {\bf 79}, 052506 (2009)

\bibitem{LabSol} L. Labzowsky and D. Solovyev, J. Phys. B: At. Mol. Opt. Phys. {\bf 37}, 3271 (2004) 

\bibitem{LSPAS} L. Labzowsky, D. Solovyev, G. Plunien, O. Andreev and G. Scedrin, J. Phys. B{\bf 40}, 525 (2007)

\bibitem{Goepp} M. G\"{o}ppert-Mayer,
Ann. Phys. (Leipzig) {\bf 9}, 273 (1931)

\bibitem{Breit} G. Breit and E. Teller,
Astrophys. J. {\bf 91}, 215 (1940)

\bibitem{Klarsfeld}
S. Klarsfeld, Phys. Lett. A {\bf 30}, 382
(1969)

\bibitem{GD82} S. P. Goldman and G. W. F. Drake, Phys.Rev. A{\bf 24}, 183 (1981)

\bibitem{PJ82} F. A. Parpia and W. R. Johnson, Phys. Rev. A{\bf 26}, 1142 (1982)

\bibitem{Santos} J.P. Santos, F. Parente, P. Indelicato Eur. Phys. J. D {\bf 3}, 43 (1998)

\bibitem{Jent} U. D. Jentschura, Phys. Rev. A{\bf 69}, 052118, (2004)

\bibitem{LSPSEur} L. N. Labzowsky, D. A. Solovyev, G. Plunien and G. Soff, Eur. Phys. J. D{\bf 37}, 335 (2006)

\bibitem{Sucher} J. Sucher, Rep. Progr. Phys. {\bf 41}, 1781 (1978)

\bibitem{SL} D. Solovyev and L. Labzowsky, Phys. Rev. A{\bf 81}, 062509 (2010)

\bibitem{SLCan} D. Solovyev and L. Labzowsky, Can. J. Phys. {\bf 89}, 123 (2011)

\bibitem{SLPSh} D. Solovyev, L. Labzowsky, G. Plunien and V. Sharipov, J. Phys. B: At. Mol. Opt. Phys. {\bf 43}, 074005 (2010)

\bibitem{answer1} S.G. Karshenboim, V.G. Ivanov and J.Chluba, arxiv:1104.486v1 [physics.atom-ph] 26 Apr. 2011

\bibitem{RZM} L. P. Rapoport, B. A. Zon and N. L. Manakov, {\it Teorija
mnogofotonnych prozessov v atomach}, ({\it Theory of the
multiphoton processes in atoms}) (Moscow, Atomizdat, 1978) (in
Russian).

\bibitem{Eur10} D. Solovyev, L. Labzowsky, A. Volotka, and G. Plunien, Eur. Phys. J. D {\bf 61}, 297 (2011)

\bibitem{Seager} S. Seager, D. Sasselov and D. Scott, ApJS {\bf 128}, 407 (2000)

\bibitem{theSpectra} L. Labzowsky, G. Klimchitskaya and Yu. Dmitriev, {\it Relativistic Effects in the Spectra of Atomic Systems}, IOP Publishing, (1993)

\end{thebibliography}
\end{document}